\documentclass[prl,aps,numerical,preprint,showkeys,longbibliography]{revtex4-1}

\usepackage{epsfig,amsmath,amssymb,txfonts,hyperref, xcolor, xr, enumitem}

\newlength{\wholefigwidth}
\newlength{\smallfigwidth}
\newlength{\halfsmallfigwidth}
\newlength{\figwidth}
\newcommand{\Fig}[1]{Fig.~\ref{#1}}

\newcommand{\Eqn}[1]{Eq.~\ref{#1}}

\makeatletter
\newcommand*{\addFileDependency}[1]{
\typeout{(#1)}
\@addtofilelist{#1}
\IfFileExists{#1}{}{\typeout{No file #1.}}
}
\makeatother

\newcommand*{\myexternaldocument}[2]{
\externaldocument[supp-]{#1/#2}
\addFileDependency{#2.tex}
\addFileDependency{#1/#2.aux}
}

\myexternaldocument{build}{Supp_mat_arxiv}

\newcommand{\FigSupp}[1]{Fig.~\ref*{supp-#1}}

\newcommand{\SecSupp}[1]{Sec.~\ref*{supp-#1}}

\newcommand{\mb}[1]{\mathbf{#1}}

\newcommand{\nnpr}{\mb{n} \rightarrow\mb{n}'}
\newcommand{\nprn}{\mb{n}' \rightarrow\mb{n}}

\newcommand{\dxa}{\delta\mb{x}^{\text{a}}_{\nnpr}}
\newcommand{\dxarev}{\delta\mb{x}^{\text{a}}_{\nprn}}

\newcommand{\dxinf}{\delta\mb{x}^{\text{a}}_{\nnpr} + \mb{y}^{\text{a}}_{\mb{n}'}-\mb{y}^{\text{a}}_{\mb{n}}}
\newcommand{\dxtil}{\delta\mb{x}^{\text{a}}_{\nnpr} + \pmb{\eta}^{\text{a}}_{\mb{n}'}-\pmb{\eta}^{\text{a}}_{\mb{n}}}

\newcommand{\ei}{ \pmb{\eta}^{\text{a}}_{\mb{n}} }
\newcommand{\ef}{ \pmb{\eta}^{\text{a}}_{\mb{n}'} }

\newcommand{\dbei}{ d_{\beta}\pmb{\eta}^{\text{a}}_{\mb{n}}  }
\newcommand{\dbef}{ d_{\beta}\pmb{\eta}^{\text{a}}_{\mb{n}'}  }

\newcommand{\tsif}{ E^{\text{TS}}_{\mb{n},\mb{n}'} }
\newcommand{\tsfi}{ E^{\text{TS}}_{\mb{n}',\mb{n}} }

\newcommand{\Qa}{Q^{\text{a}}}
\newcommand{\ddx}[2]{\delta^{(#1)}\mathbf{x}_{\nnpr}^{\text{#2}}}
\newcommand{\ddr}[2]{\delta^{(#1)}\mathbf{x}_{\nprn}^{\text{#2}}}

\newcommand{\pkin}[1]{P^{\kappa, \text{#1}}_{\mb{n},\mb{n}'}}
\newcommand{\pkinrev}[1]{P^{\kappa, \text{#1}}_{\mb{n}',\mb{n}}}

\newcommand{\vark}[1]{\Delta^{\kappa, \text{#1}}_{\tsif}}
\newcommand{\varE}{\Delta_{E_{\mb{n}}}}
\newcommand{\varG}[1]{\Delta d_{\beta}\pmb{\eta}^{\text{#1}}}

\newcommand{\thermav}[1]{\left\langle#1\right\rangle_{0}}
\newcommand{\kinav}[2]{\left\langle#1\right\rangle_{\kappa,\text{#2}}}

\newcommand{\vo}{v\textsubscript{O}}

\newcommand{\debvo}{d_{\beta}\pmb{\eta}^{ \text{\vo}} }

\begin{document}

\title{Statistical Mechanics of Thermal Diffusion in Rough Energy Landscapes}
\author{Soham Chattopadhyay}
\email{sohamc@lanl.gov, sohamc2.illinois@gmail.com}
\author{Blas P. Uberuaga}
\affiliation{Los Alamos National Laboratory}


\begin{abstract}
Starting from a variational principle, we present a broadly applicable statistical framework describing thermal diffusion in complex solids. We show that microscopic thermodynamic and kinetic fluctuations govern non-Arrhenius diffusion, with kinetic terms described by machine-learnable relative diffusion contributions. For vacancy diffusion in solid solutions, deviation from Arrheniusness may occur when ordering effects start to become important but at higher temperatures, approximately Arrhenius behavior can occur, aided by configurational entropy and competing energy fluctuations.
\end{abstract}
\maketitle

Diffusion of constituent species (particularly of mobile defects) in complex multi-component solid-state materials control several important functional properties, such as extreme-environment resistance of high entropy alloys or fast ionic transport in multi-component ionic materials \cite{Li2024, Lu2016, Lun2020, Zeng2022, Xu2023, Wang2024, Noor2026}. Accordingly, over many decades, much theoretical development (e.g., \cite{Manning1971, Murch1981, Zwanzig1988, Moleko1989, Belova2002, Nastar2007, ZauskaKotur2014, Seki2016, Allnatt2016, Thomas2020, Schuler2020, Athnes2022}) as well as computational effort (e.g., \cite{Osetsky2018, Belova2000, Xu2022_1, Manzoor2022, Reimer2025}) has been devoted to study diffusion in such complex materials. It is also well known that thermal diffusion in complex materials generally cannot be expected to be Arrhenius. However, to our knowledge, a long-standing limitation in the literature is the absence of a general fundamental theory that quantitatively and intuitively describes how the plethora of microscopic factors in these materials affect the temperature dependence of diffusion.

Recently, the variational principle for mass transport \cite{Trinkle2018}, has enabled the quantification of individual microscopic contributions to, as well as empirical non-Arrhenius formulations for diffusion \cite{Chattopadhyay2024}. Here, we utilize this theory and within the most common assumptions, develop a broadly applicable statistical mechanical framework to quantitatively describe thermally-controlled diffusion in rough energy landscapes, accounting for both microscopic thermodynamics and kinetics occurring in real complex materials.

Under near-equilibrium thermodynamics, the variational principle of mass transport gives the self-diffusion coefficient, $D^{\text{a}}$, of some species ``a'' as
\begin{align}
\nonumber D^{\text{a}} &= \inf_{\mb{y}}\frac{1}{2\text{d}} \sum_{\mb{n}, \mb{n}'}P^{0}_{\mb{n}}W_{\nnpr}\left( \dxinf \right)^2\\
&= \frac{1}{2\text{d}} \sum_{\mb{n}, \mb{n}'}P^{0}_{\mb{n}}W_{\nnpr}.\left(\dxtil\right)^2
\label{eqn_diff}
\end{align}
Here, ``d'' corresponds to the dimensions of the system, $P^{0}_{\mb{n}}$ corresponds to the equilibrium probability of state $\mb{n}$, $W_{\nnpr}$ corresponds to the rate (inverse time units) to transition from state $\mb{n}$ to state $\mb{n}'$, and $\dxa (= -\dxarev)$ corresponds to the displacement of species ``a'' during that transition. In this work, such transitions are assumed to occur via a shared transition state for both forward and backward transitions, within the framework of the widely used harmonic transition state theory (HTST) \cite{Vineyard1957}. $P^{0}_{\mb{n}}$ is accordingly defined (following \cite{Trinkle2016}) as $P^{0}_{\mb{n}} = \frac{\rho_{\mb{n}}\exp(-\beta E_{\mb{n}})}{Z}$,  where $\beta=\frac{1}{k_BT}$, $k_B$ being the Boltzmann constant and $T$ the temperature, $E_{\mb{n}}$ is the potential energy of state $\mb{n}$ and $\rho_{\mb{n}}$ is the entropic prefactor. $Z = \sum_{\mb{n}}\rho_{\mb{n}}\exp(-\beta E_{\mb{n}})$ is the equilibrium partition function. Transition rates are defined as $W_{\nnpr} = \frac{\rho_{\mb{n}, \mb{n}'}}{\rho_{\mb{n}}}\exp\left[-\beta\left( \tsif - E_{\mb{n}} \right)\right]$ and are zero for $\mb{n} = \mb{n}'$ and states between which no transition can occur. Here, $\rho_{\mb{n}, \mb{n}'} = \rho_{\mb{n}', \mb{n}}$ is an entropic prefactor of the transition state with potential energy $\tsif = \tsfi$.
Following HTST, we assume in this work that $\rho_{\mb{n}, \mb{n}'}$, $\tsif$, $\rho_{\mb{n}}$, $E_{\mb{n}}$ are temperature-independent quantities, along with $\dxa$. We recognize that generally, these quantities can also be temperature-dependent, but our goal is to examine the effects of the roughness of an energy landscape itself, the most important factor affecting diffusion in complex materials. Detailed balance is preserved, since $P^{0}_{\mb{n}} W_{\nnpr} = \frac{\rho_{\mb{n}, \mb{n}'}}{Z}\exp\left(-\beta \tsif\right) = P^{0}_{\mb{n}'}W_{\nprn}$. The ``relaxation vector,'' $\pmb{\eta}^{\text{a}}_{\mb{n}}$ of state $\mb{n}$, corresponds to the average displacement of ``a'' starting from state $\mathbf{n}$ after infinitely large number of transitions, accounting for all correlations between microscopic events (\cite{Trinkle2018}). These vectors satisfy the discrete Poisson equation:
\begin{equation}
\sum_{\mb{n}'}\Gamma_{\nnpr} (\ef - \ei) = -\sum_{\mb{n}'}\Gamma_{\nnpr} \dxa.
\label{eqn_rel_poiss}
\end{equation}
where $\Gamma_{\nnpr} = \frac{W_{\nnpr}}{\sum_{\mb{n}''}W_{\nnpr'}}$, are the transition probabilities. $\pmb{\eta}^{\text{a}}$ are thus independent of state energies $E_{\mb{n}}$.

For thermally controlled diffusion, $D^{\text{a}}$ is often macroscopically approximated with the Arrhenius form $D^{\text{a}}(T) \approx D^{\text{a}}_{T=\infty}\exp\left[-\beta Q^{\text{a}} \right]$ \cite{Allnatt_Lidiard_chp_1}. Under our assumptions, the diffusion activation barrier, $Q^{\text{a}}$, defined as the negative slope of $D^{\text{a}}$ on an Arrhenius plot, can be written from \Eqn{eqn_diff} as (derivation in supplementary \SecSupp{sec_Q})
\begin{align}
\nonumber &\Qa := -\frac{d\ln(D^{\text{a}})} {d\beta} \\
\nonumber &= \sum_{\mb{n}, \mb{n}'} \frac{ \frac{1}{2\text{d}}  P^{0}_{\mb{n}}W_{\nnpr}\left(\dxtil\right)^2}{ D^{\text{a}}  } \tsif -  \sum_{n}P^{0}_{\mb{n}}E_{\mb{n}}\\
&= \kinav{\tsif}{a} - \thermav{E_{\mb{n}}} = \kinav{\tsif - \thermav{E_{\mb{n}}} }{a}
\label{eqn_act_barr}.
\end{align}
There is an alternative expression for $\Qa$ involving $\pmb{\eta}^{\text{a}}$ in \cite{Trinkle2016}, but \Eqn{eqn_act_barr} shows that $\Qa$ is an \textit{average} (denoted as $\kinav{\ldots}{a}$) of the difference of transition state energies from the thermal average energy $\thermav{E_{\mb{n}}}$, under the purely \textit{kinetics}-controlled distribution function
\begin{equation}
\pkin{a} = \frac{ \frac{1}{2\text{d}} P^{0}_{\mb{n}}W_{\nnpr}{\left(\dxtil\right)}^2}{ D^{\text{a}}  }
\label{eqn_kin_weight},
\end{equation}
such that $\sum_{\mb{n}, \mb{n}'}\pkin{a} = 1$, $\pkin{a} = \pkinrev{a}$ and $P^{k, \text{a}}_{\mb{n},\mb{n}} = 0 \ \forall \mb{n}$. We note that $\pkin{a}$ is independent of $E_{\mb{n}}$ and hence describes the transition states independently from the equilibrium statistics of the states themselves. Also, $\pkin{a}$ includes the Boltzmann factor of a transition state (by detailed balance), but it is also scaled by the term $\left(\dxtil\right)^2$ so that it quantifies the individual relative contribution of a microscopic mechanism to \textit{net} diffusion.

\Eqn{eqn_act_barr} can also be considered a generalization of the ``superbasin transition state theory'' (SB-HTST) (see Eq. 4 in \cite{perez2014}), which models diffusion as escapes from superbasins of states with a single ``gateway'' transition state. It also shows that if $\tsif < \thermav{E_{\mb{n}}}$ and $P^{k, \text{a}}_{\mb{n},\mb{n}'} \neq 0$, such transitions contribute negative terms to $Q^{\text{a}}$.
 
In Arrhenius approximations, $\Qa$ is considered temperature-independent. However, \Eqn{eqn_act_barr} shows that the local, first-order temperature dependence of $\Qa$, in the absence of phase transformations, is given by (see supplementary \SecSupp{sec_var_rel_der} and \SecSupp{sec_dQ}),
\begin{equation}
    \frac{d\Qa} {d\beta} = -\frac{d^{2}\ln(D^{\text{a}})} {d\beta^{2}} = \varE - \vark{a} + \varG{a}
\label{eqn_act_barr_der}
\end{equation}
where,
\begin{align}
\nonumber \varE &= \thermav{E_\mb{n}^{2}} - \thermav{E_{\mb{n}}}^{2} \\
\nonumber \vark{a} &= \kinav{\left(\tsif\right)^{2}}{a} - \kinav{\tsif}{a}^{2}\\
\nonumber \varG{a} &= \frac{2}{D^{\text{a}}} \left[ \frac{1}{2\text{d}} \sum_{\mb{n}, \mb{n}'}P^{0}_{\mb{n}}W_{\nnpr}\left(\dbef -\dbei \right)^2 \right]
\end{align}
where $\dbei = \frac{d\pmb{\eta}^{\text{a}}_{\mb{n}} }{d\beta}$. From \Eqn{eqn_rel_poiss}, $\dbei$ are given by another Poisson equation (see supplementary \SecSupp{sec_var_rel_der}),
\begin{equation}
\sum_{\mb{n}'}\Gamma_{\nnpr} (\dbef - \dbei) = -\sum_{\mb{n}'} \Gamma_{\nnpr} \ddx{1}{a}
\label{eqn_rel_der_poiss}
\end{equation}
where $\ddx{1}{a} = -\tsif\left(\dxtil\right)$, with $\ddx{1}{a} = -\ddr{1}{a}$, so that $\dbei$ are also variational, satisfying \footnote{In supplementary \SecSupp{sec_general}, we show that similar variational relations also exist for relaxation derivatives with respect to macroscopic variables.}
\begin{align}
\nonumber d_{\beta}\pmb{\eta}^{\text{a}}_{\mb{n}} =\text{argmin}_{ \mathbf{y}^{\text{a}, (1)}_{\mb{n}} } \frac{1}{2\text{d}} \sum_{\mb{n}', \mb{n}''} & P^{0}_{\mb{n}'}W_{\mb{n}' \rightarrow \mb{n}''}\\
&\times \left( \delta^{(1)}\mathbf{x}^{\text{a}}_{\mb{n}' \rightarrow \mb{n}''} + \mathbf{y}^{\text{a}, (1)}_{\mb{n}''} - \mathbf{y}^{\text{a}, (1)}_{\mb{n}'} \right)^2
\label{eqn_var_rel_der}.
\end{align}
From \Eqn{eqn_rel_poiss} and \Eqn{eqn_rel_der_poiss}, $d_{\beta}\pmb{\eta}^{\text{a}}$ are independent of any constant reference state with respect to which energies are measured.

\Eqn{eqn_act_barr_der} shows that non-Arrhenius diffusion in rough energy landscapes arises from terms representing microscopic statistical fluctuations in the system, both thermodynamic and kinetic. The thermodynamic term, $\Delta_{E_\mb{n}}$, is the variance of state energies under the equilibrium Boltzmann distribution. The other two terms are kinetics-controlled. $\vark{a}$ represents the variance of the transition state energies under the distribution function $P^{k, \text{a}}_{\mb{n}\mb{n}'}$ in \Eqn{eqn_kin_weight}, while $\Delta d_{\beta}\pmb{\eta}^{\text{a}}$ contains the averaged fluctuations of $d_{\beta}\pmb{\eta}^{\text{a}}$ across connected states in the system's diffusion network.

We note that \Eqn{eqn_diff} and those that follow include all processes in general, including defect formation and annihilation. However, for simplicity, in our examples, we physically describe the terms in \Eqn{eqn_act_barr_der} by focusing on per-defect diffusion properties, setting aside the non-trivial kinetics of defect creation and annihilation.

The simple ``pure random trap'' model (see Fig. 1 in reference~\cite{Thomas2020}) explains the term $\varE$ in \Eqn{eqn_act_barr_der}. Here, all transition state energies have the same value and all transitions out of any state $\mb{n}$ have equal, temperature-independent probabilities $\Gamma_{\nnpr} = \frac{1}{z}$ ($z$ being the number of transitions out of $\mb{n}$), leading to both $\vark{a} = 0$ and $\varG{a} = 0$. A higher $\varE$ then corresponds to the system finding itself in lower energy states with decreasing temperature, with higher migration energies for diffusion, leading to increasing $\Qa$, similarly to the ``sluggish diffusion'' argument \cite{Choi2018} for high entropy systems.

Next, we look at the \textit{kinetic} terms, $\varG{a}$ and $\vark{a}$, assuming $\varE = 0$. $\varG{a}$ then accounts for the effect of kinetic correlations induced by fast, low-barrier mechanisms that that can hinder diffusion, such as repeated back and forth transitions between the same two states. As the temperature decreases, such correlation effects become stronger, making diffusion reliant on high-barrier mechanisms \cite{Chattopadhyay2026}, increasing $\Qa$. In contrast, a higher value of $\vark{a}$ indicates that both relatively high and low barrier mechanisms contribute comparably to diffusion. As the temperature decreases, to first-order, such lower barrier mechanisms start to gain more importance, reducing $\Qa$.
 \begin{figure}
     \centering
     \includegraphics[width=\smallfigwidth]{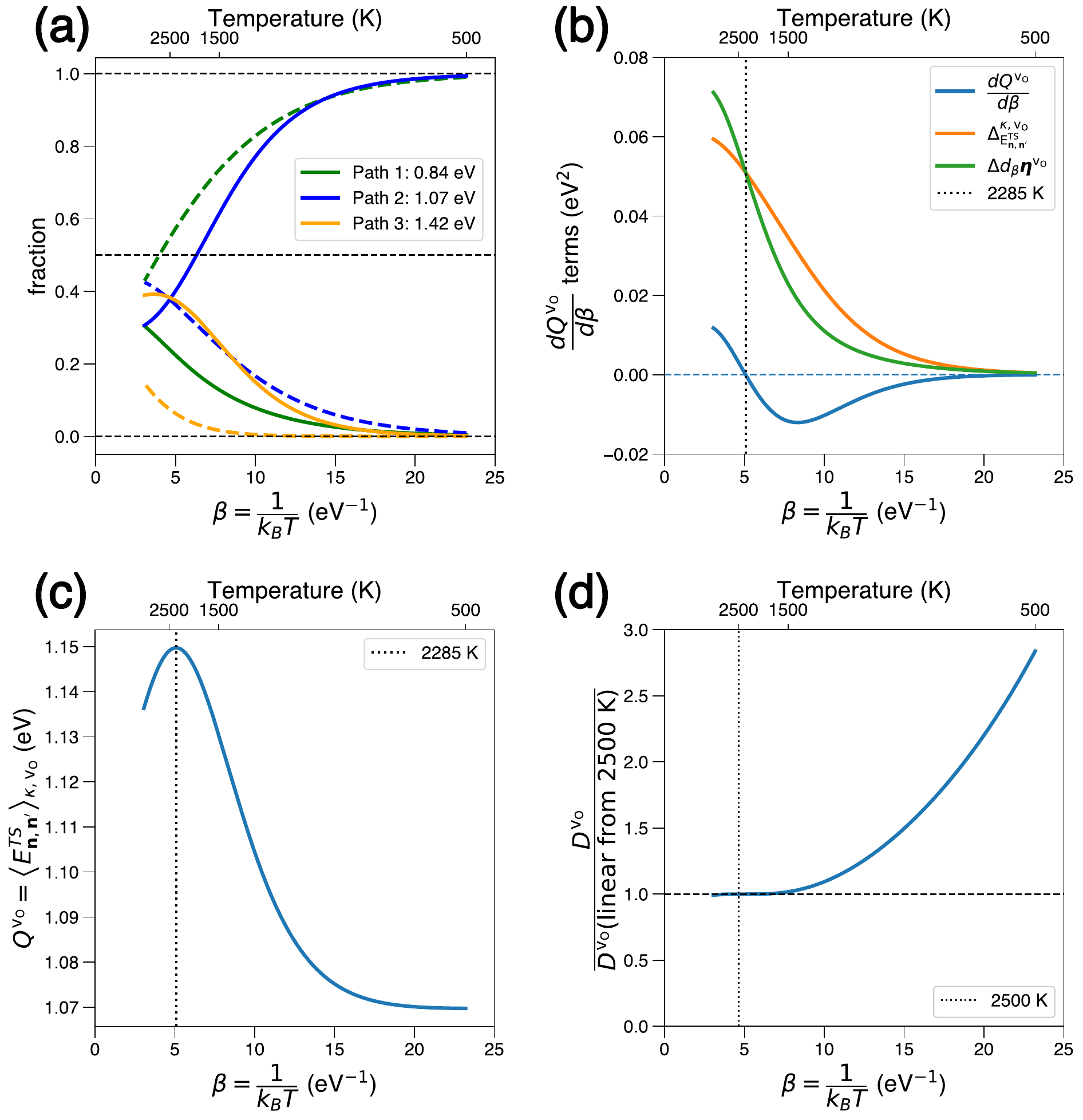}
     \caption{
     Diffusion energetics of an isolated oxygen vacancy in bixbyite Sm$_2$O$_3$. All quantities were evaluated every one K. \textbf{(a)} Fractional contribution of each mechanism class to vacancy diffusion (i.e, the sum of $\pkin{\vo}$ of all jumps belonging to a class), shown with solid lines, while dotted lines of the same color show the probability of the vacancy selecting a jump belonging to a given mechanism class from a given site. \textbf{(b)} The temperature dependence of $\frac{dQ^{ \text{v}_{\text{O}} } }{d\beta}$ along with the contributing terms in \Eqn{eqn_act_barr_der} (in this problem, $\varE = 0$). \textbf{(c)} Temperature dependence of the activation barrier $Q^{ \text{v}_{\text{O}} }$. \textbf{(d)} $D^{ \text{v}_{\text{O}} }$ at different temperatures in comparison to Arrhenius extrapolation from 2500 K. \textbf{Summary:} The saturation of jump selection probability and the relative diffusion contribution towards $1$ for ``Path 1'' and ``Path 2'' mechanisms respectively with decreasing temperature in \textbf{(a)} causes the fluctuation terms $\varG{\vo}$ and $\vark{\vo}$ to decrease in \textbf{(b)}. Overall, the competing nature of these fluctuations limits $\frac{dQ^{ \text{v}_{\text{O}} } }{d\beta}$ in \textbf{(c)}, and the deviation $D^{ \text{v}_{\text{O}} }$ in \textbf{(d)} from an Arrhenius trend.
     }

     \label{fig_Sm2O3}
 \end{figure}

To see an interplay of these kinetic terms in a realistic setting, we examine the diffusion of an isolated oxygen vacancy ({\vo}) through nearest neighbor jumps in the bixbyite-structured Sm$_2$O$_3$ compound, for which all necessary contributions in \Eqn{eqn_act_barr_der} can be computed exactly using the Green function's method \cite{Trinkle2016}. For this system, $\varE = 0$, since the oxygen sites belong to a single Wyckoff set. {\vo} in bixbyites has three main classes of migration mechanisms (see Fig. 6 in \cite{Uberuaga2015}), denoted as ``Path 1'', ``Path 2'', and ``Path 3'', for which we use migration barriers of 0.84 eV, 1.07 eV and 1.42 eV, respectively \cite{Uberuaga2026} and assume the same entropic pre-factor for all jumps. From a given state, the most probable jump is ``Path 1" but as there is only one such jump, this corresponds to a kinetic trap, since the next most probable jump will be its reverse. However, there are two symmetry-equivalent jumps each of type ``Path 2'' and ``Path 3'' that can lead to diffusion.

The diffusion activation barrier $Q^{ \text{v}_{\text{O}} }$ for $\text{v}_{\text{O}}$ and the effect of the kinetic terms $\varG{\vo}$ and $\vark{\vo}$ towards $\frac{dQ^{ \text{v}_{\text{O}} }}{d\beta}$ are shown in \Fig{fig_Sm2O3}. As the temperature decreases, the probability of the vacancy selecting the ``Path 1'' jump converges towards $1$ (\Fig{fig_Sm2O3}a, dotted lines). Thus, correlations start to saturate and $\debvo$ and hence $\varG{\vo}$ head towards zero (\Fig{fig_Sm2O3}b). On the other hand, since ``Path 1'' jumps are traps and ``Path 3'' jumps have a high barrier, with decreasing temperature, their contribution to $D^{ \text{v}_{\text{O}} }$ starts decreasing and ``Path 2'' transitions end up becoming the dominant contributors to $D^{ \text{v}_{\text{O}} }$ (\Fig{fig_Sm2O3}a, solid lines). This causes $\vark{\vo}$ to also head towards zero with decreasing temperature, with $Q^{ \text{v}_{\text{O}} }$ (\Fig{fig_Sm2O3}c) approaching the ``Path 2'' migration barrier. Also, the rate at which $\varG{\vo}$ and $\vark{\vo}$ converge towards zero are not the same, causing non-monotonic temperature-dependence of $Q^{ \text{v}_{\text{O}} }$.
The counteracting nature of $\vark{\vo}$ and $\varG{\vo}$ in \Eqn{eqn_act_barr_der}, however, limits $\frac{dQ^{ \text{v}_{\text{O}} } }{d\beta}$ (\Fig{fig_Sm2O3}b), causing a deviation in $D^{ \text{v}_{\text{O}} }$ by about a factor of 3 at 500 K from an Arrhenius extrapolation from 2500 K (\Fig{fig_Sm2O3}d).

Modern machine learning and parametric methods can be used to study non-Arrhenius effects in more complex materials \cite{Chattopadhyay2024}. As an example, we study the diffusion of a single isolated vacancy (denoted by ``v'') through nearest neighbor jumps in a Ni-20\%Cu alloy, which shows a miscibility gap at $\approx 580$ K \cite{Vrijen1978} but remains a uniform FCC solid solution at higher temperatures \cite{Iguchi2018, Turchanin2007}. We use an embedded atom potential model at \cite{Foiles1985} and the climbing image nudged elastic band method \cite{Henkelman2000} as implemented in LAMMPS \cite{LAMMPS} to compute all necessary energies. Simulation details are provided in supplementary \SecSupp{sec_CuNi}. In general, v-Cu exchanges have lower migration barriers than v-Ni exchanges, although there is appreciable overlap.

We use the ``NN + SRBC'' method described in \cite{Chattopadhyay2024} (see supplementary \SecSupp{sec_dbeta_y_SRBC} for a summary) to approximate $\pmb{\eta}^{\text{v}}_{\mb{n}}$ and $D^{\text{v}}$, along with all averages under $\pkin{v}$, from datasets of $2 \times 10^4$ vacancy jumps constructed at selected temperatures between $600-1600$ K. This procedure first uses physics-conforming neural network (NN) models (following \cite{Cohen2016} and implemented in PyTorch \cite{paszke2019}) to predict $\pmb{\eta}^{\text{v}}_{\mb{n}}$ using \Eqn{eqn_diff} from purely the atomic configuration of state $\mb{n}$. Then, a linear ``scaled residual bias'' correction (SRBC) incorporates the temperature to give improved approximations $\mb{y}^{\text{v}*}_{\mb{n}}(\text{NN + SRBC}) \approx \pmb{\eta}^{\text{v}}_{\mb{n}}$. The process effectively requires barriers to simulate two successive vacancy jumps from each starting state in a dataset. We also reused the NN model trained on a 1000 K dataset at lower temperatures, as it was found to perform well to approximate $D^{\text{v}}$ at other temperatures as well (see supplementary \FigSupp{fig_NN_evo}).

Next, a possible approximation for $d_{\beta} \pmb{\eta}^{\text{v}}$ is to directly consider $d_{\beta} \pmb{\eta}^{\text{v}} \approx d_{\beta}\mb{y}^{\text{v}*}_{\mb{n}}(\text{NN + SRBC})$. This is particularly convenient since, NN models being purely configurational, the temperature-dependence of $\mb{y}^{\text{v}*}_{\mb{n}}(\text{NN + SRBC})$ is straightforward to compute (details in supplementary \SecSupp{sec_dbeta_y_SRBC}). We call these ``direct'' approximations for $d_{\beta} \pmb{\eta}^{\text{v}}$. We can also use the variational form in \Eqn{eqn_var_rel_der} to form parametric approximations for $d_{\beta} \pmb{\eta}^{\text{v}}$. However, such approximations must be independent of the reference state energy, as per \Eqn{eqn_rel_der_poiss}. This can be achieved through a linear approximation using the same data used to optimize $\mb{y}^{\text{v}*}_{\mb{n}}(\text{NN + SRBC})$ (see supplementary \SecSupp{sec_LBAM}). We call these the ``variational'' approximations for $d_{\beta}\pmb{\eta}^{\text{v}}$. The statistics of the approximated $\pmb{\eta}^{\text{v}}$  and $d_{\beta} \pmb{\eta}^{\text{v}}$ are in supplementary \SecSupp{sec_rel_temp_sens}. Overall, both ``direct'' and ``variational'' approximations agree particularly well at high temperatures, with some relative disagreement at $800$ K, but they are also both smaller by more than an order of magnitude than $\mb{y}^{\text{v}*}_{\mb{n}}(\text{NN + SRBC})$. A maximum change of only $\approx 12.5$ \% occurs in averaged $|\mb{y}^{\text{v}*}_{\mb{n}}(\text{NN + SRBC})|$, indicating $\pmb{\eta}^{\text{v}}$ is rather temperature-insensitive within our simulated range.
\begin{figure}
     \centering
     \includegraphics[width=\smallfigwidth]{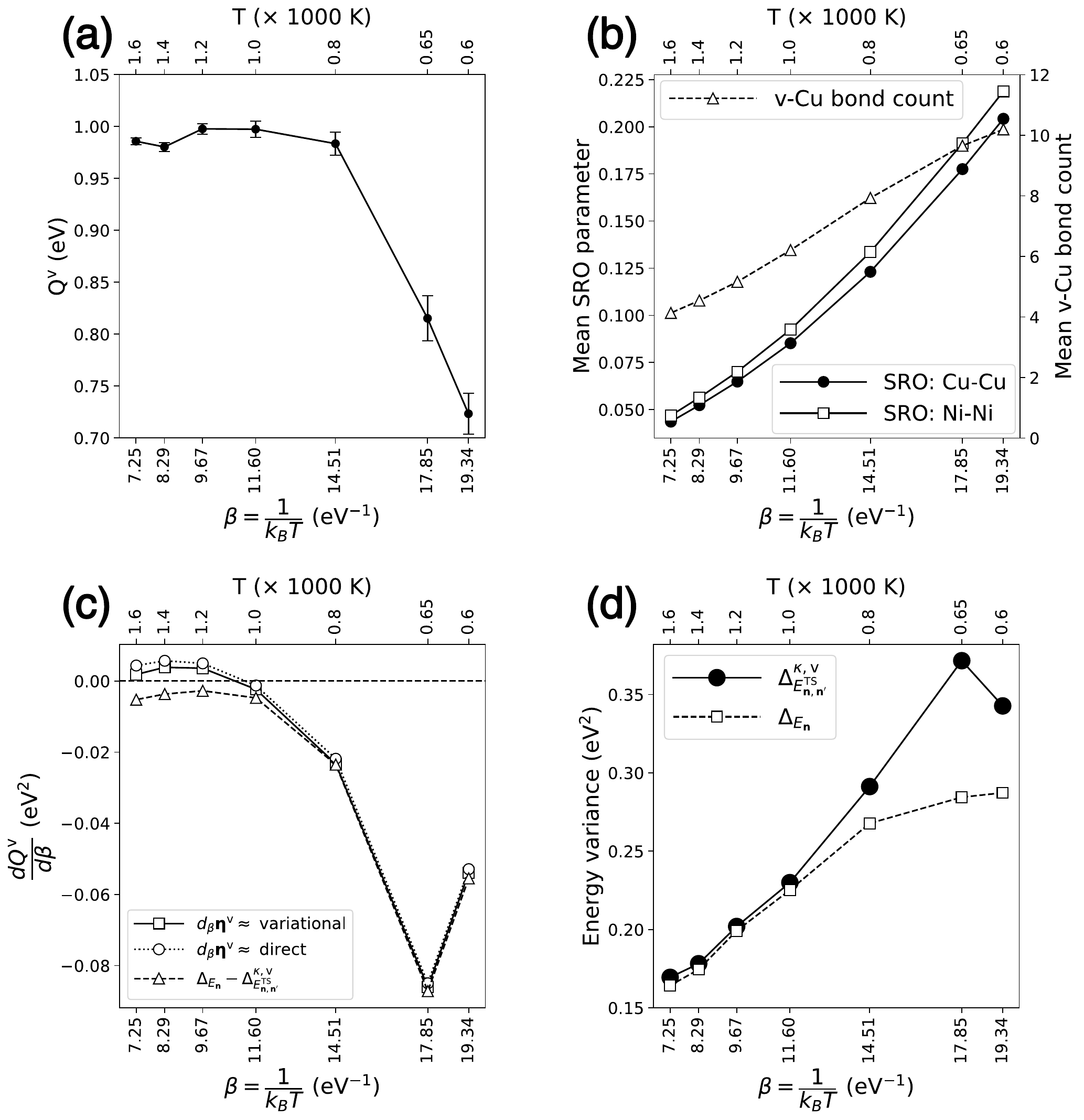}
     \caption{
     Energetics of isolated vacancy diffusion in the Ni-20\%Cu alloy. \textbf{(a)} Temperature dependence of the vacancy diffusion activation barrier $Q^{\text{v}}$. Uncertainties are shown as $\pm$ one-tenth of the standard deviation of $Q^{\text{v}}$ values computed from 100 subsets (containing 200 samples each) of the dataset of vacancy jumps at each temperature. \textbf{(b)} Temperature dependence of average short-range ordering (SRO) parameter for like atoms, as well as the average number of vacancy-Cu (v-Cu) bonds (dotted line) across equilibrium states (initial states of simulated vacancy jumps). \textbf{(c)} Temperature dependence of $\frac{dQ^{\text{v}}}{d\beta}$, with $d_{\beta}\pmb{\eta}^{\text{v}}$ computed using both the ``direct'' and ``variational'' approximations. \textbf{(d)} Temperature dependence of the energy fluctuation terms $\varE$ and $\vark{v}$ in \Eqn{eqn_act_barr_der}. \textbf{Summary:} $\varG{v}$ has only little impact on $\frac{dQ^{\text{v}}}{d\beta}$ in \textbf{(c)}. Configurational entropy causes both $\varE$ and $\vark{v}$ to decrease with increasing temperature in \textbf{(d)}, along with their difference, making $\frac{dQ^{\text{v}}}{d\beta}$ small in \textbf{(c)}, with $Q^{\text{v}}$ in \textbf{(a)} showing convergent behavior. The increased preference of Cu atom neighbors by the vacancy seen in \textbf{(b)}, along with like-atom clustering promotes vacancy diffusion via Cu-atom exchanges, leading to a prominent decrease in $Q^{\text{v}}$ in \textbf{(a)} below $800$ K. Simultaneously, non-Arrhenius effects also become important, showing larger values of $\frac{dQ^{\text{v}}}{d\beta}$ in \textbf{(c)} at lower temperatures.
 }
     \label{fig_Cu_Ni_Q_ders}
 \end{figure}

The energetics governing $D^{\text{v}}$ are shown in \Fig{fig_Cu_Ni_Q_ders}. The kinetic fluctuation term $\vark{v}$ (\Fig{fig_Cu_Ni_Q_ders}d) has an opposite trend from that of $\vark{\vo}$ in \Fig{fig_Sm2O3}, whereby, along with $\varE$, it decreases with increasing temperature. This is because $P^{0}_{\mb{n}}$ and $\pkin{v}$ become increasingly uniform as the temperature increases. Correspondingly, the energies favor values that have larger underlying microscopic configurational entropy. Furthermore, $\vark{v}$ and $\varE$ also counteract each other. As a result, vacancy diffusion in the alloy increasingly corresponds to Arrhenius behavior, with $Q^{\text{v}}$ converging towards a limiting value at high temperatures (\Fig{fig_Cu_Ni_Q_ders}a), along with lesser uncertainty in its estimates, with decreasing $|\frac{dQ^{\text{v}}}{d\beta}|$ (\Fig{fig_Cu_Ni_Q_ders}c).

Next, we see that $Q^{\text{v}}$ is substantially smaller at 650 K and 600 K than at higher temperatures (\Fig{fig_Cu_Ni_Q_ders}a). We identify three factors causing this trend. First, with decreasing temperature, Cu jumps are going to be relatively even faster than Ni jumps. Second, the vacancy increasingly prefers more Cu atom neighbors as the temperature decreases (\Fig{fig_Cu_Ni_Q_ders}b, dotted line). Third, the averaged Cu-Cu and Ni-Ni short-range order (SRO) parameters \cite{Choi2018} (\Fig{fig_Cu_Ni_Q_ders}b, solid lines) indicate increased like-atom clustering as the temperature decreases towards the miscibility gap. These factors can cause the vacancy to diffuse through Cu jumps (which generally have lower barriers), reducing $Q^{\text{v}}$. This also manifests as $\vark{v}$ growing more rapidly than $\varE$ with decreasing temperature, indicating comparable contributions to $D^{\text{v}}$ by vacancy jumps of both relative high and low transition state energies. Below $650$ K, however, $\vark{v}$ shows a decrease, along with a smaller magnitude of $\frac{dQ^{\text{v}}}{d\beta}$ at $600$ K, indicating that some mechanisms may contribute less towards $D^{\text{v}}$, possibly due to having higher barriers, similar to ``Path 3'' in \Fig{fig_Sm2O3}.
Finally, $\frac{dQ^{\text{v}}}{d\beta}$ in \Fig{fig_Cu_Ni_Q_ders} shows that the effect of $\varG{v}$ is generally small, with both ``direct'' and ``variational'' approximations for $d_{\beta}\pmb{\eta}^{\text{v}}$ showing similar results. With increasing temperatures, as $d_{\beta}\pmb{\eta}^{\text{v}}$ becomes slightly larger, $\varG{v}$ shows some growth, although its effect seems to be to only introduce changes to the sign of $\frac{dQ^{\text{v}}}{d\beta}$, where its values are generally small.
\begin{figure}
     \centering
     \includegraphics[width=0.5\smallfigwidth]{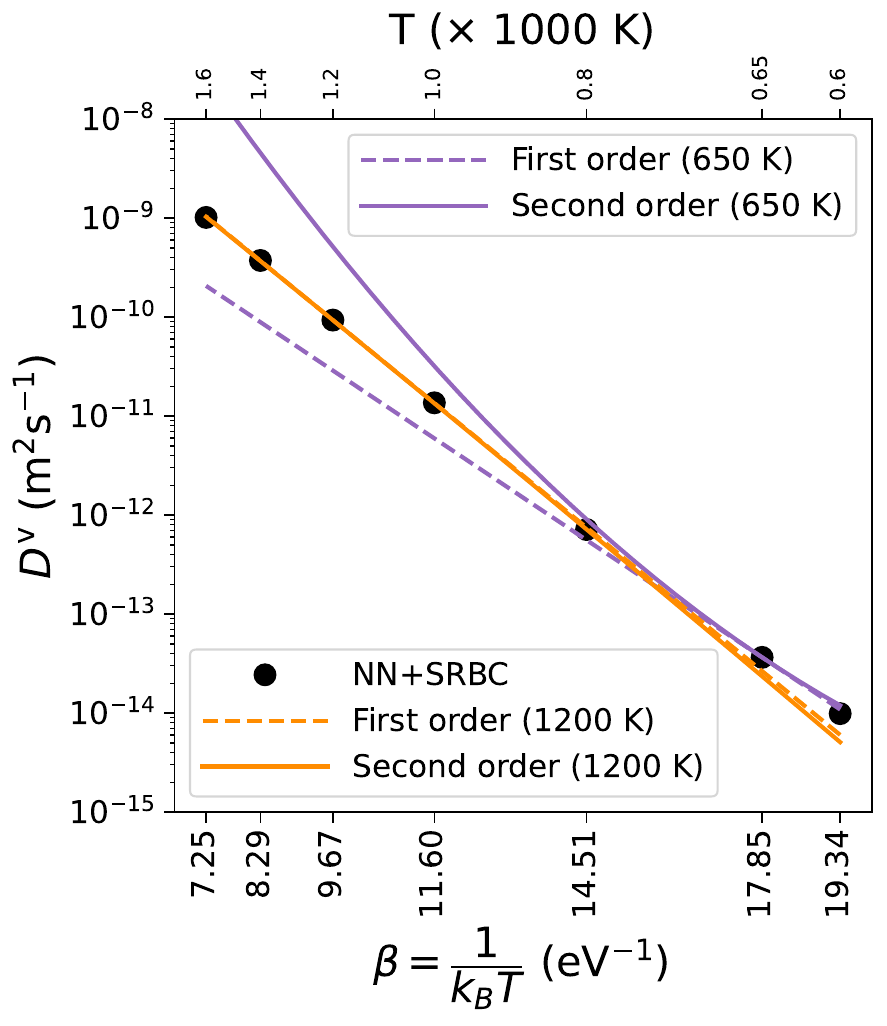}
     \caption{
     Extrapolations of $D^{\text{v}}$ from single-temperature calculations shown for 600 K and 1200 K. The ``NN+SRBC'' estimates of $D^{\text{v}}$ are indicated with filled circles. First order extrapolations are indicated with dotted lines while solid lines of the same color indicate second order extrapolations (``variational'' estimates of $\frac{dQ}{d\beta}$ we used for the second order terms) from the same temperature, both evaluated every one K. All calculations had a lattice parameter of $3.539 \AA$ and a vacancy-atom exchange attempt frequency of $10$ THz. \textbf{Summary:} At high temperatures, diffusion is nearly Arrhenius up to the limit of the stability of a solid solution, but non-Arrhenius effects become important as we approach low temperatures near the miscibility gap.
     }
     \label{fig_D_extrp}
\end{figure}

In \Fig{fig_D_extrp}, we show the second-order behavior of $D^{\text{v}}$ on an Arrhenius plot. Extrapolating $\ln(D^{\text{v}})$ from 1200 K using the computed values of $Q^{\text{v}}$ and $\frac{dQ^{\text{v}}}{d\beta}$, a first-order (or Arrhenius) extrapolation is reasonably accurate across almost the whole range of simulated temperatures, with the second order terms showing only small deviations, corresponding to small values of $\frac{dQ^{\text{v}}}{d\beta}$ at 1200 K. However, from lower temperatures, the second order terms are significant, along with different $Q^{\text{v}}$, corresponding to increasing ordering and energy fluctuations. A similar change of tracer activation barriers with decreasing temperature has also been reported experimentally elsewhere \cite{Gaertner2020} in an equiatomic FCC CoNiCrFeMn alloy.

Using the variational principle for mass transport, we derived an intuitive statistical mechanical approach that broadly describes thermal diffusion in rough energy landscapes of complex materials. Our framework quantifies competing thermodynamic and kinetic fluctuations that influence the temperature dependence of self-diffusion activation barriers. For pure materials with varying mechanisms, correlation effects can be important in governing the temperature dependence of the activation barriers. For mobile defects (a vacancy in our example) in disordered solid solutions, fluctuations of state and transition state energies under their respective distribution functions, along with configurational entropy can be more important factors, with nearly Arrhenius vacancy diffusion at higher temperatures, but significant non-Arrhenius behavior when ordering starts to become important as we approach the stability limit of a solid solution.

This work was supported by the U.S. Department of Energy, Office of Science, Basic Energy Sciences, Materials Sciences and Engineering Division (grant LANLE4BU). This research used resources provided by the Los Alamos National Laboratory Institutional Computing Program. Los Alamos National Laboratory is operated by Triad National Security, LLC, for the National Nuclear Security Administration of U.S. Department of Energy (Contract No. 89233218CNA000001).

Data is available on request. Neural network source codes are publicly available from \cite{Chattopadhyay2024}.

\bibliography{References}

@article{Vineyard1957,
  title = {Frequency factors and isotope effects in solid state rate processes},
  volume = {3},
  ISSN = {0022-3697},
  url = {http://dx.doi.org/10.1016/0022-3697(57)90059-8},
  DOI = {10.1016/0022-3697(57)90059-8},
  number = {1-2},
  journal = {Journal of Physics and Chemistry of Solids},
  publisher = {Elsevier BV},
  author = {Vineyard,  George H.},
  year = {1957},
  month = Jan,
  pages = {121–127}
}

@article{Trinkle2016,
  title = {Diffusivity and derivatives for interstitial solutes: activation energy,  volume,  and elastodiffusion tensors},
  volume = {96},
  ISSN = {1478-6443},
  url = {http://dx.doi.org/10.1080/14786435.2016.1212175},
  DOI = {10.1080/14786435.2016.1212175},
  number = {26},
  journal = {Philosophical Magazine},
  publisher = {Informa UK Limited},
  author = {Trinkle,  Dallas R.},
  year = {2016},
  month = Jul,
  pages = {2714–2735}
}

@article{Trinkle2018,
  title = {Variational Principle for Mass Transport},
  volume = {121},
  ISSN = {1079-7114},
  url = {http://dx.doi.org/10.1103/PhysRevLett.121.235901},
  DOI = {10.1103/physrevlett.121.235901},
  number = {23},
  journal = {Physical Review Letters},
  publisher = {American Physical Society (APS)},
  author = {Trinkle,  Dallas R.},
  year = {2018},
  month = dec 
}

@article{Athnes2022,
  title = {Estimating linear mass transport coefficients in solid solutions via correlation splitting and a law of total diffusion},
  volume = {6},
  ISSN = {2475-9953},
  url = {http://dx.doi.org/10.1103/PhysRevMaterials.6.013805},
  DOI = {10.1103/physrevmaterials.6.013805},
  number = {1},
  journal = {Physical Review Materials},
  publisher = {American Physical Society (APS)},
  author = {Athènes,  Manuel and Adjanor,  Gilles and Creuze,  Jér\^ome},
  year = {2022},
  month = jan 
}

@article{Schuler2020,
  title = {KineCluE: A kinetic cluster expansion code to compute transport coefficients beyond the dilute limit},
  volume = {172},
  ISSN = {0927-0256},
  url = {http://dx.doi.org/10.1016/j.commatsci.2019.109191},
  DOI = {10.1016/j.commatsci.2019.109191},
  journal = {Computational Materials Science},
  publisher = {Elsevier BV},
  author = {Schuler,  Thomas and Messina,  Luca and Nastar,  Maylise},
  year = {2020},
  month = Feb,
  pages = {109191}
}

@article{Chattopadhyay2024,
  title = {Contributions to Diffusion in Complex Materials Quantified with Machine Learning},
  volume = {132},
  ISSN = {1079-7114},
  url = {http://dx.doi.org/10.1103/PhysRevLett.132.186301},
  DOI = {10.1103/physrevlett.132.186301},
  number = {18},
  journal = {Physical Review Letters},
  publisher = {American Physical Society (APS)},
  author = {Chattopadhyay,  Soham and Trinkle,  Dallas R.},
  year = {2024},
  month = apr 
}

@article{Chattopadhyay2026,
  title = {Larger than uncorrelated vacancy diffusion contributions in chemically disordered crystalline materials},
  volume = {270},
  ISSN = {1359-6462},
  url = {http://dx.doi.org/10.1016/j.scriptamat.2025.116928},
  DOI = {10.1016/j.scriptamat.2025.116928},
  journal = {Scripta Materialia},
  publisher = {Elsevier BV},
  author = {Chattopadhyay,  Soham and Uberuaga,  Blas P.},
  year = {2026},
  month = Jan,
  pages = {116928}
}

@article{Foiles1985,
  title = {Calculation of the surface segregation of {Ni-Cu} alloys with the use of the embedded-atom method},
  volume = {32},
  ISSN = {0163-1829},
  url = {http://dx.doi.org/10.1103/PhysRevB.32.7685},
  DOI = {10.1103/physrevb.32.7685},
  number = {12},
  journal = {Physical Review B},
  publisher = {American Physical Society (APS)},
  author = {Foiles,  S. M.},
  year = {1985},
  month = Dec,
  pages = {7685–7693}
}

@article{Constantine1996,
  title = {A Multivariate Faa di Bruno Formula with Applications},
  volume = {348},
  ISSN = {0002-9947},
  url = {http://dx.doi.org/10.1090/s0002-9947-96-01501-2},
  DOI = {10.1090/s0002-9947-96-01501-2},
  number = {2},
  journal = {Transactions of the American Mathematical Society},
  publisher = {American Mathematical Society (AMS)},
  author = {Constantine,  G. and Savits,  T.},
  year = {1996},
  pages = {503–520}
}

@InProceedings{Cohen2016,
  title = 	 {Group Equivariant Convolutional Networks},
  author = 	 {Cohen, Taco and Welling, Max},
  booktitle = {Proceedings of The 33rd International Conference on Machine Learning},
  pages = 	 {2990--2999},
  year = 	 {2016},
  editor = 	 {Balcan, Maria Florina and Weinberger, Kilian Q.},
  volume = 	 {48},
  series = 	 {Proceedings of Machine Learning Research},
  address = {New York, New York, USA},
  month = 	 {20--22 Jun},
  publisher = {PMLR},
  url = {https://proceedings.mlr.press/v48/cohenc16.html}
}

@Article{LAMMPS,
  author = "A. P. Thompson and H. M. Aktulga and R. Berger and 
     D. S. Bolintineanu and W. M. Brown and P. S. Crozier and
     P. J. in 't Veld and A. Kohlmeyer and S. G. Moore and T. D. Nguyen and
     R. Shan and M. J. Stevens and J. Tranchida and C. Trott and S. J. Plimpton",
  title = "{LAMMPS} - a flexible simulation tool for
     particle-based materials modeling at the 
     atomic, meso, and continuum scales",
  journal = "Comp. Phys. Comm.",
  volume =  "271",
  pages =   "108171",
  year =    "2022",
  doi = "10.1016/j.cpc.2021.108171"
}

@article{Henkelman2000,
  title = {A climbing image nudged elastic band method for finding saddle points and minimum energy paths},
  volume = {113},
  ISSN = {1089-7690},
  url = {http://dx.doi.org/10.1063/1.1329672},
  DOI = {10.1063/1.1329672},
  number = {22},
  journal = {The Journal of Chemical Physics},
  publisher = {AIP Publishing},
  author = {Henkelman,  Graeme and Uberuaga,  Blas P. and Jónsson,  Hannes},
  year = {2000},
  month = Dec,
  pages = {9901–9904}
}

@article{EcheverriRestrepo2023,
  title = {ABC-FIRE: Accelerated Bias-Corrected Fast Inertial Relaxation Engine},
  volume = {218},
  ISSN = {0927-0256},
  url = {http://dx.doi.org/10.1016/j.commatsci.2022.111978},
  DOI = {10.1016/j.commatsci.2022.111978},
  journal = {Computational Materials Science},
  publisher = {Elsevier BV},
  author = {Echeverri Restrepo,  Sebastián and Andric,  Predrag},
  year = {2023},
  month = Feb,
  pages = {111978}
}

@inbook{Allnatt_Lidiard_chp_1,
place={Cambridge},
title={Atomic movements in solids – phenomenological equations},
booktitle={Atomic Transport in Solids},
publisher={Cambridge University Press},
author={Allnatt, A. R. and Lidiard, A. B.},
year={1993},
pages={1–49}
}

@article{Turchanin2007,
  title = {Phase equilibria and thermodynamics of binary copper systems with 3d-metals. VI. Copper-nickel system},
  volume = {46},
  ISSN = {1573-9066},
  url = {http://dx.doi.org/10.1007/s11106-007-0073-x},
  DOI = {10.1007/s11106-007-0073-x},
  number = {9-10},
  journal = {Powder Metallurgy and Metal Ceramics},
  publisher = {Springer Science and Business Media LLC},
  author = {Turchanin,  M. A. and Agraval,  P. G. and Abdulov,  A. R.},
  year = {2007},
  month = Sep,
  pages = {467–477}
}

@article{Vrijen1978,
  title = {Clustering in {Cu-Ni} alloys: A diffuse neutron-scattering study},
  volume = {17},
  ISSN = {0163-1829},
  url = {http://dx.doi.org/10.1103/PhysRevB.17.409},
  DOI = {10.1103/physrevb.17.409},
  number = {2},
  journal = {Physical Review B},
  publisher = {American Physical Society (APS)},
  author = {Vrijen,  J. and Radelaar,  S.},
  year = {1978},
  month = Jan,
  pages = {409–421}
}

@article{Iguchi2018,
  title = {On the miscibility gap of {Cu-Ni} system},
  volume = {148},
  ISSN = {1359-6454},
  url = {http://dx.doi.org/10.1016/j.actamat.2018.01.038},
  DOI = {10.1016/j.actamat.2018.01.038},
  journal = {Acta Materialia},
  publisher = {Elsevier BV},
  author = {Iguchi,  Yusuke and Katona,  Gábor L. and Cserháti,  Csaba and Langer,  Gábor A. and Erdélyi,  Zoltán},
  year = {2018},
  month = Apr,
  pages = {49–54}
}

@article{Choi2018,
  title = {Understanding the physical metallurgy of the {CoCrFeMnNi} high-entropy alloy: an atomistic simulation study},
  volume = {4},
  ISSN = {2057-3960},
  url = {http://dx.doi.org/10.1038/s41524-017-0060-9},
  DOI = {10.1038/s41524-017-0060-9},
  number = {1},
  journal = {npj Computational Materials},
  publisher = {Springer Science and Business Media LLC},
  author = {Choi,  Won-Mi and Jo,  Yong Hee and Sohn,  Seok Su and Lee,  Sunghak and Lee,  Byeong-Joo},
  year = {2018},
  month = Jan 
}

@article{HjorthLarsen2017,
  title = {The atomic simulation environment—a Python library for working with atoms},
  volume = {29},
  ISSN = {1361-648X},
  url = {http://dx.doi.org/10.1088/1361-648X/aa680e},
  DOI = {10.1088/1361-648x/aa680e},
  number = {27},
  journal = {Journal of Physics: Condensed Matter},
  publisher = {IOP Publishing},
  author = {Hjorth Larsen,  Ask and Jørgen Mortensen,  Jens and Blomqvist,  Jakob and Castelli,  Ivano E and Christensen,  Rune and Dułak,  Marcin and Friis,  Jesper and Groves,  Michael N and Hammer,  Bjørk and Hargus,  Cory and Hermes,  Eric D and Jennings,  Paul C and Bjerre Jensen,  Peter and Kermode,  James and Kitchin,  John R and Leonhard Kolsbjerg,  Esben and Kubal,  Joseph and Kaasbjerg,  Kristen and Lysgaard,  Steen and Bergmann Maronsson,  Jón and Maxson,  Tristan and Olsen,  Thomas and Pastewka,  Lars and Peterson,  Andrew and Rostgaard,  Carsten and Schiøtz,  Jakob and Sch\"{u}tt,  Ole and Strange,  Mikkel and Thygesen,  Kristian S and Vegge,  Tejs and Vilhelmsen,  Lasse and Walter,  Michael and Zeng,  Zhenhua and Jacobsen,  Karsten W},
  year = {2017},
  month = Jun,
  pages = {273002}
}

@article{Zwanzig1988,
  title = {Diffusion in a rough potential.},
  volume = {85},
  ISSN = {1091-6490},
  url = {http://dx.doi.org/10.1073/pnas.85.7.2029},
  DOI = {10.1073/pnas.85.7.2029},
  number = {7},
  journal = {Proceedings of the National Academy of Sciences},
  publisher = {National Academy of Sciences},
  author = {Zwanzig,  R},
  year = {1988},
  month = Apr,
  pages = {2029–2030}
}

@article{Seki2016,
  title = {Anomalous dimensionality dependence of diffusion in a rugged energy landscape: How pathological is one dimension?},
  volume = {144},
  ISSN = {1089-7690},
  url = {http://dx.doi.org/10.1063/1.4948936},
  DOI = {10.1063/1.4948936},
  number = {19},
  journal = {The Journal of Chemical Physics},
  publisher = {AIP Publishing},
  author = {Seki,  Kazuhiko and Bagchi,  Kaushik and Bagchi,  Biman},
  year = {2016},
  month = May 
}

@article{Thomas2020,
  title = {Vacancy diffusion in multi-principal element alloys: The role of chemical disorder in the ordered lattice},
  volume = {196},
  ISSN = {1359-6454},
  url = {http://dx.doi.org/10.1016/j.actamat.2020.06.022},
  DOI = {10.1016/j.actamat.2020.06.022},
  journal = {Acta Materialia},
  publisher = {Elsevier BV},
  author = {Thomas,  Spencer L. and Patala,  Srikanth},
  year = {2020},
  month = Sep,
  pages = {144–153}
}

@article{ZauskaKotur2014,
  title = {Variational approach to the diffusion on inhomogeneous lattices},
  volume = {304},
  ISSN = {0169-4332},
  url = {http://dx.doi.org/10.1016/j.apsusc.2013.12.133},
  DOI = {10.1016/j.apsusc.2013.12.133},
  journal = {Applied Surface Science},
  publisher = {Elsevier BV},
  author = {Załuska-Kotur,  Magdalena A.},
  year = {2014},
  month = Jun,
  pages = {122–126}
}

@article{Manning1971,
  title = {Correlation Factors for Diffusion in Nondilute Alloys},
  volume = {4},
  ISSN = {0556-2805},
  url = {http://dx.doi.org/10.1103/PhysRevB.4.1111},
  DOI = {10.1103/physrevb.4.1111},
  number = {4},
  journal = {Physical Review B},
  publisher = {American Physical Society (APS)},
  author = {Manning,  John R.},
  year = {1971},
  month = Aug,
  pages = {1111–1121}
}

@article{Murch1981,
  title = {Diffusion,  correlation,  and percolation in a random alloy},
  volume = {43},
  ISSN = {1460-6992},
  url = {http://dx.doi.org/10.1080/01418618108239403},
  DOI = {10.1080/01418618108239403},
  number = {1},
  journal = {Philosophical Magazine A},
  publisher = {Informa UK Limited},
  author = {Murch,  G. E. and Rothman,  S. J.},
  year = {1981},
  month = Jan,
  pages = {229–238}
}

@article{Moleko1989,
  title = {A self-consistent theory of matter transport in a random lattice gas and some simulation results},
  volume = {59},
  ISSN = {1460-6992},
  url = {http://dx.doi.org/10.1080/01418618908220335},
  DOI = {10.1080/01418618908220335},
  number = {1},
  journal = {Philosophical Magazine A},
  publisher = {Informa UK Limited},
  author = {Moleko,  L. K. and Allnatt,  A. R. and Allnatt,  E. L.},
  year = {1989},
  month = Jan,
  pages = {141–160}
}

@article{Belova2002,
  title = {Collective and tracer diffusion kinetics in the ternary random alloy},
  volume = {14},
  ISSN = {0953-8984},
  url = {http://dx.doi.org/10.1088/0953-8984/14/28/301},
  DOI = {10.1088/0953-8984/14/28/301},
  number = {28},
  journal = {Journal of Physics: Condensed Matter},
  publisher = {IOP Publishing},
  author = {Belova,  I V and Allnatt,  A R and Murch,  G E},
  year = {2002},
  month = Jul,
  pages = {6897–6907}
}

@article{Allnatt2016,
  title = {A high accuracy diffusion kinetics formalism for random multicomponent alloys: application to high entropy alloys},
  volume = {96},
  ISSN = {1478-6443},
  url = {http://dx.doi.org/10.1080/14786435.2016.1219785},
  DOI = {10.1080/14786435.2016.1219785},
  number = {28},
  journal = {Philosophical Magazine},
  publisher = {Informa UK Limited},
  author = {Allnatt,  A. R. and Paul,  T. R. and Belova,  I. V. and Murch,  G. E.},
  year = {2016},
  month = Aug,
  pages = {2969–2985}
}

@article{Nastar2007,
  title = {A self-consistent mean field theory for diffusion in alloys},
  volume = {134},
  ISSN = {1364-5498},
  url = {http://dx.doi.org/10.1039/B605834E},
  DOI = {10.1039/b605834e},
  journal = {Faraday Discuss.},
  publisher = {Royal Society of Chemistry (RSC)},
  author = {Nastar,  Maylise and Barbe,  Vincent},
  year = {2007},
  pages = {331–342}
}

@article{Gaertner2020,
  title = {Tracer diffusion in single crystalline {CoCrFeNi} and {CoCrFeMnNi} high-entropy alloys: Kinetic hints towards a low-temperature phase instability of the solid-solution?},
  volume = {187},
  ISSN = {1359-6462},
  url = {http://dx.doi.org/10.1016/j.scriptamat.2020.05.060},
  DOI = {10.1016/j.scriptamat.2020.05.060},
  journal = {Scripta Materialia},
  publisher = {Elsevier BV},
  author = {Gaertner,  Daniel and Kottke,  Josua and Chumlyakov,  Yury and Hergem\"{o}ller,  Fabian and Wilde,  Gerhard and Divinski,  Sergiy V.},
  year = {2020},
  month = Oct,
  pages = {57–62}
}

@misc{kingma2017,
      title={Adam: A Method for Stochastic Optimization}, 
      author={Diederik P. Kingma and Jimmy Ba},
      year={2017},
      eprint={1412.6980},
      archivePrefix={arXiv},
      primaryClass={cs.LG},
      url={https://arxiv.org/abs/1412.6980}, 
}

@inproceedings{paszke2019,
  title     = {PyTorch: An Imperative Style, High-Performance Deep Learning Library},
  author    = {Paszke, A. and Gross, S. and Massa, F. and ... and Chintala, S.},
  booktitle = {Advances in Neural Information Processing Systems 32},
  pages     = {8024--8035},
  year      = {2019},
  url       = {https://papers.nips.cc/paper/9015-pytorch-an-imperative-style-high-performance-deep-learning-library}
}

@article{Osetsky2018,
  title = {On the existence and origin of sluggish diffusion in chemically disordered concentrated alloys},
  volume = {22},
  ISSN = {1359-0286},
  url = {http://dx.doi.org/10.1016/j.cossms.2018.05.003},
  DOI = {10.1016/j.cossms.2018.05.003},
  number = {3},
  journal = {Current Opinion in Solid State and Materials Science},
  publisher = {Elsevier BV},
  author = {Osetsky,  Yuri N. and Béland,  Laurent K. and Barashev,  Alexander V. and Zhang,  Yanwen},
  year = {2018},
  month = jun,
  pages = {65–74}
}

@article{Belova2000,
  title = {Collective diffusion in the binary random alloy},
  volume = {80},
  ISSN = {1460-6992},
  url = {http://dx.doi.org/10.1080/01418610008212070},
  DOI = {10.1080/01418610008212070},
  number = {3},
  journal = {Philosophical Magazine A},
  publisher = {Informa UK Limited},
  author = {Belova,  I. V. and Murch,  G. E.},
  year = {2000},
  month = Mar,
  pages = {599–607}
}

@article{Xu2022_1,
  doi = {10.1016/j.actamat.2022.118051},
  url = {https://doi.org/10.1016/j.actamat.2022.118051},
  year = {2022},
  month = aug,
  publisher = {Elsevier {BV}},
  volume = {234},
  pages = {118051},
  author = {Biao Xu and Jun Zhang and Shihua Ma and Yaoxu Xiong and Shasha Huang and J.J. Kai and Shijun Zhao},
  title = {Revealing the crucial role of rough energy landscape on self-diffusion in high-entropy alloys based on machine learning and kinetic {Monte Carlo}},
  journal = {Acta Materialia}
}

@article{Manzoor2022,
  title = {Influence of Defect Thermodynamics on Self-Diffusion in Complex Concentrated Alloys with Chemical Ordering},
  volume = {74},
  ISSN = {1543-1851},
  url = {http://dx.doi.org/10.1007/s11837-022-05477-9},
  DOI = {10.1007/s11837-022-05477-9},
  number = {11},
  journal = {JOM},
  publisher = {Springer Science and Business Media LLC},
  author = {Manzoor,  Anus and Zhang,  Yongfeng},
  year = {2022},
  month = sep,
  pages = {4107–4120}
}

@article{Reimer2025,
  title = {Prediction of vacancy defect diffusion paths in high entropy alloys via machine learning on molecular dynamics data},
  volume = {138},
  ISSN = {1089-7550},
  url = {http://dx.doi.org/10.1063/5.0280842},
  DOI = {10.1063/5.0280842},
  number = {7},
  journal = {Journal of Applied Physics},
  publisher = {AIP Publishing},
  author = {Reimer,  C. and Saidi,  P. and Casert,  C. and Beeler,  C. and Tetsassi Feugmo,  C. G. and Whitelam,  S. and Mansouri,  E. and Martinez,  A. and Beland,  L. and Tamblyn,  I.},
  year = {2025},
  month = Aug 
}

@article{Zeng2022,
  title = {High-entropy mechanism to boost ionic conductivity},
  volume = {378},
  ISSN = {1095-9203},
  url = {http://dx.doi.org/10.1126/science.abq1346},
  DOI = {10.1126/science.abq1346},
  number = {6626},
  journal = {Science},
  publisher = {American Association for the Advancement of Science (AAAS)},
  author = {Zeng,  Yan and Ouyang,  Bin and Liu,  Jue and Byeon,  Young-Woon and Cai,  Zijian and Miara,  Lincoln J. and Wang,  Yan and Ceder,  Gerbrand},
  year = {2022},
  month = Dec,
  pages = {1320–1324}
}

@article{Noor2026,
  title = {Configurational Entropy-Driven Oxide-Ion Transport in Rare-Earth High-Entropy Fluorite Electrolytes for Low-Temperature Fuel Cells},
  volume = {14},
  ISSN = {2168-0485},
  url = {http://dx.doi.org/10.1021/acssuschemeng.6c00158},
  DOI = {10.1021/acssuschemeng.6c00158},
  number = {17},
  journal = {ACS Sustainable Chemistry \& Engineering},
  publisher = {American Chemical Society (ACS)},
  author = {Noor,  Asma and Guo,  Chunyu and Yousaf,  Muhammad and Xiuxiu,  Li and Bibi,  Bushra and Hayat,  Qaisar and Lu,  Yuzheng},
  year = {2026},
  month = Apr,
  pages = {8259–8270}
}

@article{Wang2024,
  title = {High‐Entropy Strategy Flattening Lithium Ion Migration Energy Landscape to Enhance the Conductivity of Garnet‐Type Solid‐State Electrolytes},
  volume = {35},
  ISSN = {1616-3028},
  url = {http://dx.doi.org/10.1002/adfm.202416389},
  DOI = {10.1002/adfm.202416389},
  number = {9},
  journal = {Advanced Functional Materials},
  publisher = {Wiley},
  author = {Wang,  Shuhan and Wen,  Xiaojuan and Huang,  Zhenweican and Xu,  Haoyang and Fan,  Fengxia and Wang,  Xinxiang and Tian,  Guilei and Liu,  Sheng and Liu,  Pengfei and Wang,  Chuan and Zeng,  Chenrui and Shu,  Chaozhu and Liang,  Zhenxing},
  year = {2024},
  month = Nov 
}

@article{Xu2023,
  title = {High-entropy electrolytes in boosting battery performance},
  volume = {2},
  ISSN = {2752-5724},
  url = {http://dx.doi.org/10.1088/2752-5724/ace8ab},
  DOI = {10.1088/2752-5724/ace8ab},
  number = {4},
  journal = {Materials Futures},
  publisher = {IOP Publishing},
  author = {Xu,  Jijian},
  year = {2023},
  month = Aug,
  pages = {047501}
}

@article{Lun2020,
  title = {Cation-disordered rocksalt-type high-entropy cathodes for {Li}-ion batteries},
  volume = {20},
  ISSN = {1476-4660},
  url = {http://dx.doi.org/10.1038/s41563-020-00816-0},
  DOI = {10.1038/s41563-020-00816-0},
  number = {2},
  journal = {Nature Materials},
  publisher = {Springer Science and Business Media LLC},
  author = {Lun,  Zhengyan and Ouyang,  Bin and Kwon,  Deok-Hwang and Ha,  Yang and Foley,  Emily E. and Huang,  Tzu-Yang and Cai,  Zijian and Kim,  Hyunchul and Balasubramanian,  Mahalingam and Sun,  Yingzhi and Huang,  Jianping and Tian,  Yaosen and Kim,  Haegyeom and McCloskey,  Bryan D. and Yang,  Wanli and Clément,  Raphaële J. and Ji,  Huiwen and Ceder,  Gerbrand},
  year = {2020},
  month = Oct,
  pages = {214–221}
}

@article{Li2024,
  title = {Tunable interstitial and vacancy diffusivity by chemical ordering control in {CrCoNi} medium-entropy alloy},
  volume = {10},
  ISSN = {2057-3960},
  url = {http://dx.doi.org/10.1038/s41524-024-01322-6},
  DOI = {10.1038/s41524-024-01322-6},
  number = {1},
  journal = {npj Computational Materials},
  publisher = {Springer Science and Business Media LLC},
  author = {Li,  Yangen and Du,  Jun-Ping and Shinzato,  Shuhei and Ogata,  Shigenobu},
  year = {2024},
  month = Jun 
}

@article{Lu2016,
  title = {Enhancing radiation tolerance by controlling defect mobility and migration pathways in multicomponent single-phase alloys},
  volume = {7},
  ISSN = {2041-1723},
  url = {http://dx.doi.org/10.1038/ncomms13564},
  DOI = {10.1038/ncomms13564},
  number = {1},
  journal = {Nature Communications},
  publisher = {Springer Science and Business Media LLC},
  author = {Lu,  Chenyang and Niu,  Liangliang and Chen,  Nanjun and Jin,  Ke and Yang,  Taini and Xiu,  Pengyuan and Zhang,  Yanwen and Gao,  Fei and Bei,  Hongbin and Shi,  Shi and He,  Mo-Rigen and Robertson,  Ian M. and Weber,  William J. and Wang,  Lumin},
  year = {2016},
  month = Dec 
}

@article{Uberuaga2015,
  title = {Interpreting oxygen vacancy migration mechanisms in oxides using the layered structure motif},
  volume = {103},
  ISSN = {0927-0256},
  url = {http://dx.doi.org/10.1016/j.commatsci.2014.10.013},
  DOI = {10.1016/j.commatsci.2014.10.013},
  journal = {Computational Materials Science},
  publisher = {Elsevier BV},
  author = {Uberuaga,  Blas Pedro and Sickafus,  Kurt E.},
  year = {2015},
  month = Jun,
  pages = {216–223}
}

@article{Uberuaga2026,
  title = {The impact of chemistry on anion migration in bixbyite-structured lanthanide oxides},
  volume = {444},
  ISSN = {0167-2738},
  url = {http://dx.doi.org/10.1016/j.ssi.2026.117264},
  DOI = {10.1016/j.ssi.2026.117264},
  journal = {Solid State Ionics},
  publisher = {Elsevier BV},
  author = {Uberuaga,  Blas Pedro and Chattopadhyay,  Soham},
  year = {2026},
  month = Oct,
  pages = {117264}
}

@article{Perez2014,
  title = {Diffusion and transformation kinetics of small helium clusters in bulk tungsten},
  volume = {90},
  ISSN = {1550-235X},
  url = {http://dx.doi.org/10.1103/PhysRevB.90.014102},
  DOI = {10.1103/physrevb.90.014102},
  number = {1},
  journal = {Physical Review B},
  publisher = {American Physical Society (APS)},
  author = {Perez,  Danny and Vogel,  Thomas and Uberuaga,  Blas P.},
  year = {2014},
  month = Jul 
}

\end{document}


\title{Statistical Mechanics of Thermal Diffusion in Rough Energy Landscapes - Supplementary Material}

\author{Soham Chattopadhyay}
\author{Blas P. Uberuaga}
\affiliation{Los Alamos National Laboratory}

\date{\today}

\begin{abstract}
Supplementary Material containing the following. \textbf{(Section S1)}: derivations of the equations for the self-diffusion activation barrier and its first derivative with respect to temperature and approximations for the diffusion relaxation vector derivatives with respect to inverse temperature. \textbf{(Section S2)}: Details of Monte Carlo and kinetic Monte Carlo simulations to generate datasets for vacancy diffusion. Also contains details about training neural network models for computing vacancy diffusion coefficients, and statistics of approximated vacancy relaxation vectors and their derivatives. (\textbf{Section S3}): General expressions for derivatives of diffusion relaxation vectors.
\end{abstract}
\maketitle

\section{Derivations}
\label{sec_dervs}
\subsection{Diffusion Activation Barrier}
\label{sec_Q}
We start with the variational principle of mass transport \cite{Trinkle2018}, which gives the self-diffusion coefficient of a species ``a'' as
\begin{equation}
D^{\text{a}} = \inf_{\mb{y}}\frac{1}{2\text{d}} \sum_{\mb{n}, \mb{n}'}P^{0}_{\mb{n}}W_{\nnpr}\left( \dxinf \right)^2 = \frac{1}{2\text{d}} \sum_{\mb{n}, \mb{n}'}P^{0}_{\mb{n}}W_{\nnpr}.\left(\dxtil\right)^2
\label{eqn_var_pri}
\end{equation}
The quantities in this equation are defined in the main text, but we repeat some definitions here for completeness. Namely, the state probabilities $P^{0}_{\mb{n}}$ are,
\begin{equation}
P^{0}_{\mb{n}} = \frac{\rho_{\mb{n}}\exp(-\beta E_{\mb{n}})}{Z}
\label{eqn_prob}
\end{equation}
where $\beta=\frac{1}{k_BT}$, $k_B$ being the Boltzmann constant and $T$ the temperature, $Z = \sum_{\mb{n}}\rho_{\mb{n}}\exp(-\beta E_{\mb{n}})$ is the equilibrium partition function, $E_{\mb{n}}$ is the energy of state $\mb{n}$, and $\rho_{\mb{n}}$ is the entropic pre-factor.
The transition rates, $W_{\nnpr}$ are taken to be non-zero only when $\mb{n} \neq \mb{n}'$ and there exists a transition from $\mb{n}$ to $\mb{n}'$, and zero otherwise, and are defined as \cite{Vineyard1957, Trinkle2016}
\begin{equation}
W_{\nnpr} = \frac{\rho_{\mb{n}, \mb{n}'}}{\rho_{\mb{n}}}\exp\left[-\beta\left( \tsif - E_{\mb{n}} \right)\right]
\label{eqn_rate}.
\end{equation}
Here,
$\rho_{\mb{n}, \mb{n}'} = \rho_{\mb{n}', \mb{n}}$ is the entropic prefactor of the transition state between states $\mb{n}$ and $\mb{n}'$ with energy $\tsif = \tsfi$ (note that the rate has inverse time units). We assume that $\rho_{\mb{n}, \mb{n}'}$, $\tsif$, $\rho_{\mb{n}}$, $E_{\mb{n}}$ and $\dxa$ are all temperature-independent quantities. \Eqn{eqn_prob} and \Eqn{eqn_rate} also preserve detailed balance, such that
\begin{equation}
P^{0}_{\mb{n}} W_{\nnpr} = P^{0}_{\mb{n}'}W_{\nprn} = \frac{\rho_{\mb{n}, \mb{n}'}}{Z}\exp\left(-\beta \tsif\right)
\label{eqn_det_bal}.
\end{equation}

The diffusion activation barrier for species ``a'' is defined as
\begin{equation}
\Qa := -\frac{d\ln(\Da)} {d\beta} = -\frac{1}{\Da}\frac{d\Da}{d\beta}
\label{eqn_Q_main}
\end{equation}
To evaluate $\Qa$, we use \Eqn{eqn_var_pri} to get
\begin{equation}
\frac{d\Da}{d\beta} = \frac{1}{2\dm} \sum_{\mb{n}, \mb{n}'} \tsder\left(\dxtil\right)^2 + \frac{1}{2\dm} \sum_{\mb{n}, \mb{n}'} P^{0}_{\mb{n}}W_{\nnpr}\kinder
\label{eqn_Dder_1}
\end{equation}

We first note that the second term in \Eqn{eqn_Dder_1} vanishes, since
\begin{align}
\frac{1}{2}\nonumber&\sum_{\mb{n}, \mb{n}'} P^{0}_{\mb{n}}W_{\nnpr}\kinder = \sum_{\mb{n}, \mb{n}'} P^{0}_{\mb{n}}W_{\nnpr}\left(\dxtil\right).\left( \dbetanp - \dbetan \right)\\
\nonumber&= \sum_{\mb{n}, \mb{n}'} P^{0}_{\mb{n}}W_{\nnpr}\left(\dxtil\right).\dbetanp - \sum_{\mb{n}, \mb{n}'} P^{0}_{\mb{n}}W_{\nnpr}\left(\dxtil\right).\dbetan\\
&= \sum_{\mb{n}, \mb{n}'} P^{0}_{\mb{n}'}W_{\nprn}\left(\dxtil\right).\dbetanp - \sum_{\mb{n}, \mb{n}'} P^{0}_{\mb{n}}W_{\nnpr}\left(\dxtil\right).\dbetan \label{eqn_use_det_bal} \\
\nonumber&= -\sum_{\mb{n}, \mb{n}'} P^{0}_{\mb{n}'}W_{\nprn}\left(\dxtilrev\right).\dbetanp - \sum_{\mb{n}, \mb{n}'} P^{0}_{\mb{n}}W_{\nnpr}\left(\dxtil\right).\dbetan\\
&= -\sum_{\mb{n}, \mb{n}'} P^{0}_{\mb{n}}W_{\nnpr}\left(\dxtil\right).\dbetan - \sum_{\mb{n}, \mb{n}'} P^{0}_{\mb{n}}W_{\nnpr}\left(\dxtil\right).\dbetan \label{eqn_ex_inds} \\
\nonumber&= -2\sum_{\mb{n}} P^{0}_{\mb{n}}\dbetan \sum_{\mb{n}'} W_{\nnpr}\left(\dxtil\right) = \mb{0}.
\end{align}
where in the step \Eqn{eqn_use_det_bal}, we use detailed balance in the first term, and in the step \Eqn{eqn_ex_inds}, we exchange the indices $\mb{n}$ and $\mb{n}'$ in the first term since they run over all states. In the last step, $\sum_{\mb{n}'} W_{\nnpr}\left(\dxtil\right) = \mb{0}$.
We then have from \Eqn{eqn_Dder_1}
\begin{equation}
\frac{d\Da}{d\beta} = \frac{1}{2\dm} \sum_{\mb{n}, \mb{n}'} \tsder\left(\dxtil\right)^2
\label{eqn_Dder_2}
\end{equation}
Now, using \Eqn{eqn_det_bal} and \Eqn{eqn_prob}, we can write
\begin{align}
&\nonumber\tsder = \frac{d}{d\beta}\left[\frac{\rho_{\mb{n}, \mb{n}'}}{Z}\exp\left(-\beta \tsif\right)\right]\\
\nonumber&=\rho_{\mb{n}, \mb{n}'}\frac{ Z\frac{d \exp\left(-\beta \tsif\right) }{d\beta} - \exp\left(-\beta \tsif\right)\frac{dZ}{d\beta} }{Z^2}\\
\nonumber&= -\tsif\frac{\rho_{\mb{n}, \mb{n}'}}{Z}\exp\left(-\beta \tsif\right) - \frac{\rho_{\mb{n}, \mb{n}'}}{Z}\exp\left(-\beta \tsif\right)\frac{\frac{d}{d\beta}\left[ \sum_{\mb{n}}\rho_{\mb{n}}\exp(-\beta E_{\mb{n}} ) \right]}{Z}\\
\nonumber&= -\tsif\frac{\rho_{\mb{n}, \mb{n}'}}{Z}\exp\left(-\beta \tsif\right) + \frac{\rho_{\mb{n}, \mb{n}'}}{Z}\exp\left(-\beta \tsif\right)\left(\sum_{\mb{n}} E_{\mb{n}}\frac{\rho_{\mb{n}}\exp(-\beta E_{\mb{n}} )}{Z}\right)\\
\nonumber&= -\frac{\rho_{\mb{n}, \mb{n}'}}{Z}\exp\left(-\beta \tsif\right)\left( \tsif - \thermav{E_{\mb{n}}} \right)\\
&= -P^{0}_{\mb{n}}W_{\nnpr}\left( \tsif - \thermav{E_{\mb{n}}} \right) \label{eqn_det_bal_der}
\end{align}
where $\thermav{E_{\mb{n}}} = \sum_{\mb{n}}P^{0}_{\mb{n}}E_{\mb{n}}$. Putting this result in \Eqn{eqn_Dder_2} and using \Eqn{eqn_var_pri} gives
\begin{align}
&\nonumber\frac{d\Da}{d\beta} = -\frac{1}{2\dm} \sum_{\mb{n}, \mb{n}'} P^{0}_{\mb{n}}W_{\nnpr}\tsif\left(\dxtil\right)^2 + \thermav{E_{\mb{n}}}\frac{1}{2\dm} \sum_{\mb{n}, \mb{n}'} P^{0}_{\mb{n}}W_{\nnpr}\left(\dxtil\right)^2\\
&=-\frac{1}{2\dm} \sum_{\mb{n}, \mb{n}'} P^{0}_{\mb{n}}W_{\nnpr}\tsif\left(\dxtil\right)^2 + \thermav{E_{\mb{n}}}\Da
\end{align}
so that from \Eqn{eqn_Q_main}
\begin{align}
\nonumber&\Qa = -\frac{1}{\Da}\frac{d\Da}{d\beta} = \sum_{\mb{n}, \mb{n}'} \frac{\frac{1}{2\dm} P^{0}_{\mb{n}}W_{\nnpr}\left(\dxtil\right)^2}{\Da}\tsif - \thermav{E_{\mb{n}}}\\
&=\sum_{\mb{n},\mb{n}'}\pkin{a}\tsif - \thermav{E_{\mb{n}}} = \kinav{\tsif}{a} - \thermav{E_{\mb{n}}} \label{eqn_Q_avgs}
\end{align}
which is the expression for $\Qa$ in our main article.
and defines the weights of the microscopic mechanisms for the diffusion of species ``a'' as
\begin{equation}
\pkin{a} = \frac{\frac{1}{2\dm} P^{0}_{\mb{n}}W_{\nnpr}\left(\dxtil\right)^2}{\Da} = \pkinrev{a}
\label{eqn_p_kin}
\end{equation}

We note that averages of the form $\kinav{\ldots}{a}$, can be empirically computed by using single-step kinetic Monte Carlo trajectories (following \cite{Chattopadhyay2024}) to evaluate both the numerator and denominator terms, since $\sum_{\mb{n}, \mb{n}'} P^{0}_{\mb{n}}W_{\nnpr}(\ldots) = \sum_{\mb{n}, \mb{n}'} P^{0}_{\mb{n}}\Gamma_{\nnpr}\tau_{\mb{n}}^{-1}(\ldots)$, where $\tau_{\mb{n}}^{-1} = \sum_{\mb{n}'}W_{\nnpr}$ is the total escape rate from a state and $\Gamma_{\nnpr} = \frac{W_{\nnpr}}{\sum_{\mb{n}'}W_{\nnpr}}$ is the relative probability of choosing a transition from a given state.

\subsection{Variational Principle for Relaxation Vector Derivative with Respect to Inverse Temperature}
\label{sec_var_rel_der}
The relaxation vectors $\pmb{\eta}^{\text{a}}$ satisfy
\begin{equation}
\sum_{\mb{n}'}W_{\nnpr}(\ef - \ei) = -\sum_{\mb{n}'}W_{\nnpr}\dxa
\label{eqn_rel_poiss}
\end{equation}
Or, using \Eqn{eqn_rate}, we get
\begin{equation}
\sum_{\mb{n}'}\frac{\rho_{\mb{n}, \mb{n}'}}{\rho_{\mb{n}}}\exp\left[-\beta\left( \tsif - E_{\mb{n}} \right)\right](\ef - \ei) = -\sum_{\mb{n}'} \frac{\rho_{\mb{n}, \mb{n}'}}{\rho_{\mb{n}}}\exp\left[-\beta\left( \tsif - E_{\mb{n}} \right)\right] \dxa
\end{equation}
Or,
\begin{equation}
\sum_{\mb{n}'} \rho_{\mb{n}, \mb{n}'} \exp\left(-\beta \tsif\right) (\ef - \ei) = -\sum_{\mb{n}'} \rho_{\mb{n}, \mb{n}'} \exp\left(-\beta \tsif \right) \dxa
\end{equation}
Differentiating both sides with respect to $\beta$, we get
\begin{align}
\nonumber -\sum_{\mb{n}'}  \tsif \rho_{\mb{n}, \mb{n}'}\exp\left(-\beta \tsif\right) (\ef - \ei) + \sum_{\mb{n}'} \rho_{\mb{n}, \mb{n}'}\exp\left(-\beta \tsif\right) \left(\dbetanp - \dbetan\right) =\\
\sum_{\mb{n}'} \tsif \rho_{\mb{n}, \mb{n}'} \exp\left(-\beta \tsif \right) \dxa
\end{align}
Or,
\begin{align}
\sum_{\mb{n}'} \rho_{\mb{n}, \mb{n}'}\exp\left(-\beta \tsif\right) \left(\dbetanp - \dbetan\right) =
\sum_{\mb{n}'} \rho_{\mb{n}, \mb{n}'}\exp\left(-\beta \tsif \right)\left[\tsif \left( \dxa + \ef - \ei \right)\right]
\end{align}
Denoting
\begin{equation}
\ddx{1}{a} = -\tsif \left( \dxa + \ef - \ei \right)
\label{eqn_Del_x},
\end{equation}
we can write
\begin{align}
\sum_{\mb{n}'} \rho_{\mb{n}, \mb{n}'}\exp\left(-\beta \tsif\right) \left(\dbetanp - \dbetan\right) =
-\sum_{\mb{n}'} \rho_{\mb{n}, \mb{n}'}\exp\left(-\beta \tsif \right)\ddx{1}{a}
\end{align}
Or, multiplying both sides by $\frac{\exp(\beta E_{\mb{n}})}{\rho_{\mb{n}}}$,
\begin{equation}
\sum_{\mb{n}'} W_{\nnpr} \left(\dbetanp - \dbetan\right) =
-\sum_{\mb{n}'} W_{\nnpr}\ddx{1}{a}
\label{eqn_rel_der_poiss}
\end{equation}
which is similar to \Eqn{eqn_rel_poiss}, with $\ddx{1}{a} = -\ddr{1}{a} $.

Similar to \Eqn{eqn_var_pri} for $\ei$, the derivatives can therefore be obtained variationaly as
\begin{equation}
\dbetan = \text{argmin}_{\mathbf{y}^{(1)}_{\mb{n}}} \frac{1}{2\dm} \sum_{\mb{n'},\mb{n}''}P^{0}_{\mb{n'}}W_{\mb{n}' \rightarrow \mb{n}''}\left( \delta^{1}\mathbf{x}^{\text{a}}_{\mb{n}' \rightarrow \mb{n}''} + \mathbf{y}^{(1)}_{\mb{n}''} - \mathbf{y}^{(1)}_{\mb{n}'}  \right)^2
\label{eqn_var_rel_der}
\end{equation}
Since $\sum_{\mb{n}'} W_{\nnpr}\left(\dxtil\right) = \mb{0}$, $\dbetan$ are invariant to any constant reference with respect to which all transition state energies are measured.

\subsection{First Derivative of Diffusion Activation Barrier}
\label{sec_dQ}
To compute $\frac{d\Qa}{d\beta}$, we first compute (using \Eqn{eqn_p_kin})
\begin{align}
&\frac{d\pkin{a}}{d\beta} = \nonumber\frac{1}{2\dm}\frac{d}{d\beta}\left( \frac{P^{0}_{\mb{n}}W_{\nnpr}\left(\dxtil\right)^2}{\Da} \right)\\
&\nonumber=\frac{
    \Da \frac{d}{d\beta} \left[P^{0}_{\mb{n}}W_{\nnpr}\left(\dxtil\right)^2\right] - P^{0}_{\mb{n}}W_{\nnpr}\left(\dxtil\right)^2 \frac{d\Da}{d\beta}
}{2\dm(\Da)^2}\\
&=
\frac{
\ddb\left[P^{0}_{\mb{n}}W_{\nnpr}\left(\dxtil\right)^2\right]
}{2\dm\Da} -
\left(\frac{\frac{1}{2\dm}P^{0}_{\mb{n}}W_{\nnpr}\left(\dxtil\right)^2}{\Da}\right)\left(\frac{1}{\Da}\frac{d\Da}{d\beta}\right)
\end{align}
Using \Eqn{eqn_Q_main} and \Eqn{eqn_p_kin} in the second term, we write
\begin{equation}
\frac{d\pkin{a}}{d\beta} =
\frac{
\ddb\left[P^{0}_{\mb{n}}W_{\nnpr}\left(\dxtil\right)^2\right]
}{2\dm\Da} + \pkin{a}\Qa
\label{eqn_d_p_kin_main}
\end{equation}
Then, we compute the derivative in the first term on the right-hand side of \Eqn{eqn_d_p_kin_main},
\begin{align}
&\nonumber\ddb\left[P^{0}_{\mb{n}}W_{\nnpr}\left(\dxtil\right)^2\right]\\
&\nonumber=\frac{d\left(P^{0}_{\mb{n}}W_{\nnpr}\right)}{d\beta}\left(\dxtil\right)^2 + P^{0}_{\mb{n}}W_{\nnpr}\ddb\left(\dxtil\right)^2\\
\nonumber&=\frac{d\left(P^{0}_{\mb{n}}W_{\nnpr}\right)}{d\beta}\left(\dxtil\right)^2 + 2 P^{0}_{\mb{n}}W_{\nnpr}\left(\dxtil\right).\left( \dbetanp - \dbetan \right) \\
&= -P^{0}_{\mb{n}}W_{\nnpr}\left(\dxtil\right)^2\left( \tsif - \thermav{E_{\mb{n}}} \right) + 2 P^{0}_{\mb{n}}W_{\nnpr}\left(\dxtil\right).\left( \dbetanp - \dbetan \right)
\label{eqn_d_kin}
\end{align}
where we used \Eqn{eqn_det_bal_der}. Putting \Eqn{eqn_d_kin} in \Eqn{eqn_d_p_kin_main}, we get
\begin{align}
\nonumber\frac{d\pkin{a}}{d\beta} &= -\frac{P^{0}_{\mb{n}}W_{\nnpr}\left(\dxtil\right)^2\left( \tsif - \thermav{E_{\mb{n}}} \right)}{2\dm\Da}\\
\nonumber &+\frac{2P^{0}_{\mb{n}}W_{\nnpr}\left(\dxtil\right).\left( \dbetanp - \dbetan \right)}{2\dm\Da}\\ &+\pkin{a}\Qa
\end{align}
Or, using \Eqn{eqn_p_kin}
\begin{align}
\nonumber\frac{d\pkin{a}}{d\beta} &= -\pkin{a}\left( \tsif - \thermav{E_{\mb{n}}} \right) + \pkin{a}\Qa +\frac{P^{0}_{\mb{n}}W_{\nnpr}\left(\dxtil\right).\left( \dbetanp - \dbetan \right)}{\dm\Da}\\ &
\label{eqn_d_p_kin}
\end{align}
Now, we see from \Eqn{eqn_Q_avgs} that
\begin{equation}
\frac{d\Qa}{d\beta} = \sum_{\mb{n}, \mb{n}'}\tsif\frac{d\pkin{a}}{d\beta} - \frac{d\thermav{E_{\mb{n}}}}{d\beta} = \sum_{\mb{n}, \mb{n}'}\tsif\frac{d\pkin{a}}{d\beta} + \varE
\label{eqn_dQ_main}
\end{equation}
where $\frac{d\thermav{E_{\mb{n}}}}{d\beta} = -\thermav{E_{\mb{n}}^2} + \thermav{E_{\mb{n}}}^2 = - \varE$ is the negative variance of state energies.
Using \Eqn{eqn_d_p_kin}, we can then write,
\begin{align}
\nonumber\frac{d\Qa}{d\beta} &= -\sum_{\mb{n}, \mb{n}'}\pkin{a}\tsif\left( \tsif - \thermav{E_{\mb{n}}} \right) + \Qa\sum_{\mb{n}, \mb{n}'}\pkin{a}\tsif\\
&\nonumber +\sum_{\mb{n}, \mb{n}'} \frac{P^{0}_{\mb{n}}\tsif W_{\nnpr}\left(\dxtil\right).\left( \dbetanp - \dbetan \right)}{\dm\Da}\\
& + \varE
\end{align}
Or, using \Eqn{eqn_Q_avgs} for $\Qa$
\begin{align}
\nonumber\frac{d\Qa}{d\beta} &=  -\sum_{\mb{n}, \mb{n}'}\pkin{a}\tsif\left( \tsif - \thermav{E_{\mb{n}}} \right) + \left( \sum_{\mb{n},\mb{n}'}\pkin{a}\tsif - \thermav{E_{\mb{n}}} \right)\sum_{\mb{n}, \mb{n}'}\pkin{a}\tsif\\
&\nonumber +\sum_{\mb{n}, \mb{n}'} \frac{P^{0}_{\mb{n}}\tsif W_{\nnpr}\left(\dxtil\right).\left( \dbetanp - \dbetan \right)}{\dm\Da}\\
& + \varE
\end{align}
Or,
\begin{align}
\nonumber\frac{d\Qa}{d\beta} &=
-\sum_{\mb{n}, \mb{n}'}\pkin{a}\left(\tsif\right)^2 + \thermav{E_{\mb{n}}} \sum_{\mb{n}, \mb{n}'}\pkin{a}\tsif
+ \left( \sum_{\mb{n},\mb{n}'}\pkin{a}\tsif\right)^2 - \thermav{E_{\mb{n}}} \nonumber\sum_{\mb{n}, \mb{n}'}\pkin{a}\tsif\\
&\nonumber +\sum_{\mb{n}, \mb{n}'} \frac{P^{0}_{\mb{n}}\tsif W_{\nnpr}\left(\dxtil\right).\left( \dbetanp - \dbetan \right)}{\dm\Da}\\
& + \varE
\end{align}
Or,
\begin{equation}
\frac{d\Qa}{d\beta} = \frac{1}{\dm\Da}\sum_{\mb{n}, \mb{n}'} P^{0}_{\mb{n}}\tsif W_{\nnpr}\left(\dxtil\right).\left( \dbetanp - \dbetan \right) + \varE - \vark{a}
\label{eqn_dQ_1}
\end{equation}
where, $\vark{a} = \sum_{\mb{n}, \mb{n}'}\pkin{a}\left(\tsif\right)^2 - \left( \sum_{\mb{n},\mb{n}'}\pkin{a}\tsif \right)^2$ is the variance of the transition states energies under the kinetic distribution function $\pkin{a}$.

Now,
\begin{align}
 \nonumber&\sum_{\mb{n}, \mb{n}'} P^{0}_{\mb{n}}\tsif W_{\nnpr}\left(\dxtil\right).\left( \dbetanp - \dbetan \right) \\
 \nonumber&= \sum_{\mb{n}, \mb{n}'} P^{0}_{\mb{n}}\tsif W_{\nnpr}\left(\dxtil\right).\dbetanp - \sum_{\mb{n}, \mb{n}'} P^{0}_{\mb{n}}\tsif W_{\nnpr}\left(\dxtil\right).\dbetan\\
 &= \sum_{\mb{n}, \mb{n}'} P^{0}_{\mb{n}'}\tsfi W_{\nprn}\left(\dxtil\right).\dbetanp - \sum_{\mb{n}, \mb{n}'} P^{0}_{\mb{n}}\tsif W_{\nnpr}\left(\dxtil\right).\dbetan \label{use_det_bal_and_symm} \\
\nonumber&= -\sum_{\mb{n}, \mb{n}'} P^{0}_{\mb{n}'}\tsfi W_{\nprn}\left(\dxtilrev\right).\dbetanp - \sum_{\mb{n}, \mb{n}'} P^{0}_{\mb{n}}\tsif W_{\nnpr}\left(\dxtil\right).\dbetan\\
\nonumber&= -\sum_{\mb{n}, \mb{n}'} P^{0}_{\mb{n}}\tsif W_{\nnpr}\left(\dxtil\right).\dbetan - \sum_{\mb{n}, \mb{n}'} P^{0}_{\mb{n}}\tsif W_{\nnpr}\left(\dxtil\right).\dbetan\\
\nonumber&= 2\sum_{\mb{n}} P^{0}_{\mb{n}}\dbetan.\sum_{\mb{n}'}W_{\nnpr}\left[ -\tsif\left(\dxtil\right) \right]\\
 &= 2\sum_{\mb{n}} P^{0}_{\mb{n}}\dbetan.\sum_{\mb{n}'}W_{\nnpr}\left(\ddx{1}{a}\right)
\end{align}
where, in step \Eqn{use_det_bal_and_symm}, in the first term, we used detailed balance $P^{0}_{\mb{n}}W_{\nnpr} = P^{0}_{\mb{n}'}W_{\nprn}$ and $\tsif = \tsfi$. In the last step, we used the definition from \Eqn{eqn_Del_x}. Using equation \Eqn{eqn_rel_der_poiss} in the equation above, we can then write
\begin{align}
\nonumber&\sum_{\mb{n}, \mb{n}'} P^{0}_{\mb{n}}\tsif W_{\nnpr}\left(\dxtil\right).\left( \dbetanp - \dbetan \right)\\
\nonumber&= -2\sum_{\mb{n},\mb{n}'} P^{0}_{\mb{n}}W_{\nnpr}\dbetan.\left( \dbetanp - \dbetan \right)\\
\nonumber&= -\sum_{\mb{n},\mb{n}'} P^{0}_{\mb{n}}W_{\nnpr}\dbetan.\left( \dbetanp - \dbetan \right) - \sum_{\mb{n},\mb{n}'} P^{0}_{\mb{n}}W_{\nnpr}\dbetan.\left( \dbetanp - \dbetan \right)\\
\nonumber&= \sum_{\mb{n},\mb{n}'} P^{0}_{\mb{n}'}W_{\nprn}\dbetan.\left( \dbetan - \dbetanp \right) - \sum_{\mb{n},\mb{n}'} P^{0}_{\mb{n}}W_{\nnpr}\dbetan.\left( \dbetanp - \dbetan \right)\\
\nonumber&= \sum_{\mb{n},\mb{n}'} P^{0}_{\mb{n}}W_{\nnpr}\dbetanp.\left( \dbetanp - \dbetan \right) - \sum_{\mb{n},\mb{n}'} P^{0}_{\mb{n}}W_{\nnpr}\dbetan.\left( \dbetanp - \dbetan \right)\\
&=\sum_{\mb{n},\mb{n}'} P^{0}_{\mb{n}}W_{\nnpr}\left( \dbetanp - \dbetan \right)^2 \label{eqn_drel_fluc}
\end{align}

Using \Eqn{eqn_drel_fluc} in \Eqn{eqn_dQ_1}, we get
\begin{equation}
\frac{d\Qa}{d\beta} = \frac{1}{\dm\Da}\sum_{\mb{n},\mb{n}'} P^{0}_{\mb{n}}W_{\nnpr}\left( \dbetanp - \dbetan \right)^2 + \varE - \vark{a}
\label{eqn_dQ_2}
\end{equation}
Or,
\begin{equation}
\frac{d\Qa}{d\beta} = \frac{2}{\Da}\left[\frac{1}{2\dm}\sum_{\mb{n},\mb{n}'} P^{0}_{\mb{n}}W_{\nnpr}\left( \dbetanp - \dbetan \right)^2\right] + \varE - \vark{a}
\label{eqn_dQ}
\end{equation}
which is the equation used in the main paper for $\frac{d\Qa}{d\beta}$.

\subsection{Derivatives of NN + SRBC approximations}
\label{sec_dbeta_y_SRBC}
For complex systems, where relaxation vectors $\pmb{\eta}^{\text{a}}_{\mb{n}}$ are parametrically approximated using the NN + SRBC method as described in \cite{Chattopadhyay2024}, a simple way to approximate $\frac{d\pmb{\eta}^{\text{a}}_{\mb{n}}}{d\beta}$ is to directly consider $\frac{d\pmb{\eta}^{\text{a}}_{\mb{n}}}{d\beta} \approx \frac{d \mathbf{y}^{\text{a}*}_{\mb{n}}\text{(NN + SRBC)}}{d\beta}$. We recall from \cite{Chattopadhyay2024} that
\begin{equation}
\mb{y}^{\text{a}*}_{\mb{n}}(\text{NN + SRBC}) = \mb{y}^{\text{a}*}_{\mb{n}'}(\text{NN}) + \sum_{i} \theta^{i*}_{\text{SRBC}} \mb{b}^{\text{a}}_{i}({\mb{n}})
\label{eqn_NN_SRBC}
\end{equation}
Here, $\mb{y}^{\text{a}*}(\text{NN}) \approx \pmb{\eta}^{\text{a}}_{\mb{n}}$ are optimal neural network predictions for $\pmb{\eta}^{\text{a}}$ which are obtained by optimizing \Eqn{eqn_var_pri} and are purely configurational, temperature-independent functions. These are then improved on using the equation above to give $\mb{y}^{\text{a}*}_{\mb{n}}(\text{NN + SRBC}) \approx \pmb{\eta}^{\text{a}}_{\mb{n}}$. The sum over $i$ runs over all the species (both defect and atomic species). The basis functions for each species are defined as
\begin{equation}
\mb{b}^{\text{a}}_{i}({\mb{n}}) = 
\begin{cases}
\sum_{\mb{n}'}\Gamma_{\nnpr} \left[ \dxf{a} + \mb{y}^{\text{a}*}_{\mb{n}'}(\text{NN}) - \mb{y}^{\text{a}*}_{\mb{n}}(\text{NN}) \right] \ \text{if } {i} = \text{a} \\
\sum_{\mb{n}'} \Gamma_{\nnpr} \left(\dxf{i} \right) \ \text{otherwise}
\end{cases}
\label{eqn_basis_srbc}
\end{equation}
where 
\begin{equation}
\Gamma_{\nnpr} = \frac{W_{\nnpr}}{\sum_{\mb{n}'} W_{\nnpr} }
\label{eqn_j_prob}
\end{equation}
is the probability of a transition from an initial state $\mb{n}$.

Using the approximations \Eqn{eqn_NN_SRBC} in  \Eqn{eqn_var_pri}, the linear expansion coefficients $\theta^{i*}_{\text{SRBC}}$ in \Eqn{eqn_NN_SRBC}, are obtained by solving the system of equations
\footnote{In Eq. S4 of \cite{Chattopadhyay2024}, due to a typographical error, a negative sign is missing on the right-hand side. \Eqn{eqn_LBAM_SRBC} is the correct equation for solving the optimal linear expansion coefficients and also correctly reflects all code implementation and results in both \cite{Chattopadhyay2024} and the current work.}
\begin{equation}
\sum_{j}\overline{W}_{ij} \theta^{j*}_{\text{SRBC}} = - \overline{b}_{i}
\label{eqn_LBAM_SRBC}
\end{equation}
where both $i$ and $j$ run over the species and
\begin{align}
\overline{W}_{ij} &= \sum_{\mb{n}, \mb{n}'} P^{0}_{\mb{n}} W_{\nnpr} \Delta \mb{b}^{\text{a}}_{i}(\nnpr).\Delta \mb{b}^{\text{a}}_{j}(\nnpr) = \overline{W}_{ji} \label{eqn_LBAM_terms_Wbar}\\
\overline{b}_{i} &= \sum_{\mb{n}, \mb{n}'} P^{0}_{\mb{n}} W_{\nnpr} \left[ \delta\mathbf{x}^{\text{a}}_{\mb{n} \rightarrow \mb{n}'} + \mb{y}^{\text{a}*}_{\mb{n}'}(\text{NN}) - \mb{y}^{\text{a}*}_{\mb{n}}(\text{NN})  \right].\Delta \mb{b}^{\text{a}}_{i}(\nnpr) \label{eqn_LBAM_terms_bbar}
\end{align}
where $\Delta \mb{b}^{\text{a}}_{i}(\nnpr) = \mb{b}^{\text{a}}_{i}(\mb{n}') - \mb{b}^{\text{a}}_{i}(\mb{n})$.

From \Eqn{eqn_NN_SRBC},
\begin{equation}
\frac{d \mathbf{y}^{\text{a}*}_{\mb{n}}\text{(NN + SRBC)}}{d\beta} = \sum_{i} \left( \theta^{i*}_{\text{SRBC}} \frac{d\mb{b}^{\text{a}}_{i}({\mb{n}})}{d\beta} + \frac{d\theta^{i*}_{\text{SRBC}}}{d\beta} \mb{b}^{\text{a}}_{i}({\mb{n}}) \right)
\end{equation}

From \Eqn{eqn_basis_srbc},
\begin{equation}
\frac{d\mb{b}^{\text{a}}_{i}({\mb{n}})}{d\beta} = 
\begin{cases}
\sum_{\mb{n}'}\frac{d\Gamma_{\nnpr}}{d\beta} \left[ \dxf{a} + \mb{y}^{\text{a}*}_{\mb{n}'}(\text{NN}) - \mb{y}^{\text{a}*}_{\mb{n}}(\text{NN}) \right] \ \text{if } {i} = \text{a} \\
\sum_{\mb{n}'} \frac{d\Gamma_{\nnpr}}{d\beta} \left(\dxf{i} \right) \ \text{otherwise}
\end{cases}
\label{eqn_dbeta_basis_srbc}
\end{equation}
where, from \Eqn{eqn_j_prob} and \Eqn{eqn_rate} we find
\begin{equation}
\frac{d\Gamma_{\nnpr}}{d\beta} = \Gamma_{\nnpr} \sum_{\mb{n}''}\Gamma_{\nnpr'} \left( E^{\text{TS}}_{\mb{n},\mb{n}''} - E^{\text{TS}}_{\mb{n},\mb{n}'}  \right)
\label{eqn_d_gamma}
\end{equation}

From \Eqn{eqn_LBAM_SRBC}, we find that the derivatives $\frac{d\theta^{j*}_{\text{SRBC}}}{d\beta}$ can be found by solving
\begin{equation}
\sum_{j}\overline{W}_{ij} \frac{d\theta^{j*}_{\text{SRBC}}}{d\beta} = - \frac{d\overline{b}_{i}}{d\beta} - \sum_{j} \frac{d\overline{W}_{ij} }{d\beta} \theta^{j*}_{\text{SRBC}}
\end{equation}

where, from \Eqn{eqn_LBAM_terms_Wbar} and \Eqn{eqn_LBAM_terms_bbar}, we have
\begin{align}
\nonumber \frac{d\overline{W}_{ij}}{d\beta} = \frac{d\overline{W}_{ji}}{d\beta} = &\sum_{\mb{n}, \mb{n}'} \frac{dP^{0}_{\mb{n}} W_{\nnpr}}{d\beta} \Delta \mb{b}^{\text{a}}_{i}(\nnpr).\Delta \mb{b}^{\text{a}}_{j}(\nnpr) \\
& + \sum_{\mb{n}, \mb{n}'} P^{0}_{\mb{n}} W_{\nnpr}\Delta \mb{b}^{\text{a}}_{i}(\nnpr). \frac{d\Delta \mb{b}^{\text{a}}_{j}(\nnpr)}{d\beta}\\
& + \sum_{\mb{n}, \mb{n}'} P^{0}_{\mb{n}} W_{\nnpr} \frac{d\Delta \mb{b}^{\text{a}}_{i}(\nnpr)}{d\beta}.\Delta \mb{b}^{\text{a}}_{j}(\nnpr)
\end{align}
and
\begin{align}
\frac{d\overline{b}_{i}}{d\beta} = &\sum_{\mb{n}, \mb{n}'} \frac{d P^{0}_{\mb{n}} W_{\nnpr} }{d\beta} \left[ \delta\mathbf{x}^{\text{a}}_{\mb{n} \rightarrow \mb{n}'} + \mb{y}^{\text{a}*}_{\mb{n}'}(\text{NN}) - \mb{y}^{\text{a}*}_{\mb{n}}(\text{NN})  \right].\Delta \mb{b}^{\text{a}}_{i}(\nnpr) \\
& + \sum_{\mb{n}, \mb{n}'} P^{0}_{\mb{n}} W_{\nnpr} \left[ \delta\mathbf{x}^{\text{a}}_{\mb{n} \rightarrow \mb{n}'} + \mb{y}^{\text{a}*}_{\mb{n}'}(\text{NN}) - \mb{y}^{\text{a}*}_{\mb{n}}(\text{NN})  \right] . \frac{d \Delta \mb{b}^{\text{a}}_{i}(\nnpr) }{d\beta}
\end{align}
From \Eqn{eqn_det_bal_der}, we can rewrite these as
\begin{align}
\nonumber \frac{d\overline{W}_{ij}}{d\beta} = \frac{d\overline{W}_{ji}}{d\beta} = & \sum_{\mb{n}, \mb{n}'} P^{0}_{\mb{n}} W_{\nnpr}\Delta \mb{b}^{\text{a}}_{i}(\nnpr). \frac{d\Delta \mb{b}^{\text{a}}_{j}(\nnpr)}{d\beta}\\
&+ \sum_{\mb{n}, \mb{n}'} P^{0}_{\mb{n}} W_{\nnpr} \frac{d\Delta \mb{b}^{\text{a}}_{i}(\nnpr)}{d\beta}.\Delta \mb{b}^{\text{a}}_{j}(\nnpr)\\
&- \sum_{\mb{n}, \mb{n}'} P^{0}_{\mb{n}}W_{\nnpr}\left( \tsif - \thermav{E_{\mb{n}}} \right) \Delta \mb{b}^{\text{a}}_{i}(\nnpr).\Delta \mb{b}^{\text{a}}_{j}(\nnpr)
\end{align}
and
\begin{align}
\frac{d\overline{b}_{i}}{d\beta} = &\sum_{\mb{n}, \mb{n}'} P^{0}_{\mb{n}} W_{\nnpr} \left[ \delta\mathbf{x}^{\text{a}}_{\mb{n} \rightarrow \mb{n}'} + \mb{y}^{\text{a}*}_{\mb{n}'}(\text{NN}) - \mb{y}^{\text{a}*}_{\mb{n}}(\text{NN})  \right] . \frac{d \Delta \mb{b}^{\text{a}}_{i}(\nnpr) }{d\beta}\\
&- \sum_{\mb{n}, \mb{n}'} P^{0}_{\mb{n}}W_{\nnpr}\left( \tsif - \thermav{E_{\mb{n}}} \right)  \left[ \delta\mathbf{x}^{\text{a}}_{\mb{n} \rightarrow \mb{n}'} + \mb{y}^{\text{a}*}_{\mb{n}'}(\text{NN}) - \mb{y}^{\text{a}*}_{\mb{n}}(\text{NN})  \right].\Delta \mb{b}^{\text{a}}_{i}(\nnpr)
\end{align}
and use \Eqn{eqn_dbeta_basis_srbc} for the rest of the terms.

\subsection{Linear Basis Approximation for Relaxation Vector Derivatives}
\label{sec_LBAM}
We can also use \Eqn{eqn_var_rel_der} to variationally approximate for $\frac{d\pmb{\eta}^{\text{a}}_{\mb{n}}}{d\beta}$. The process is similar to the SRBC method in \cite{Chattopadhyay2024} and is derived from the general ``Linear Basis Approximation Method'' \cite{Trinkle2018}. We first expand the relaxation vector derivatives $\frac{d\pmb{\eta}^{\text{a}}_{\mb{n}}}{d\beta}$ using the expansion:
\begin{equation}
\mathbf{y}^{\text{a}, (1)}_{\mb{n}} = \sum_{i}\theta^{i} \mb{b}^{\text{a}}_{i}({\mb{n}}) \approx \frac{d\pmb{\eta}^{\text{a}}_{\mb{n}}}{d\beta},
\label{eqn_LBAM_dy}
\end{equation}
where the sum over $i$ runs over all species and $\mb{b}^{\text{a}}_{i}({\mb{n}})$ are the SRBC  basis functions in \Eqn{eqn_basis_srbc}

Following \Eqn{eqn_var_rel_der}, the optimal values, $\theta^{i*}$, of the coefficients $\theta^{i}$, are then obtained by solving,
\begin{equation}
\sum_{\beta}\overline{W}^{(1)}_{ij} \theta^{j*} = -\overline{b}^{(1)}_{i}
\label{eqn_lbam_dder}
\end{equation}
where the sum over $j$ also runs over all species and
\begin{align}
\overline{W}^{(1)}_{ij}  &= \sum_{\mb{n}, \mb{n}'} P^{0}_{\mb{n}} W_{\nnpr} \Delta \mb{b}^{\text{a}}_{i}(\nnpr).\Delta \mb{b}^{\text{a}}_{j}(\nnpr) = \overline{W}^{(1)}_{ji}\\
\overline{b}^{(1)}_{i} &= \sum_{\mb{n}, \mb{n}'} P^{0}_{\mb{n}} W_{\nnpr} \delta^{(1)}\mathbf{x}^{\text{a}}_{\mb{n} \rightarrow \mb{n}'} . \Delta \mb{b}^{\text{a}}_{i}(\nnpr)
\end{align}
$\delta^{(1)}\mathbf{x}^{\text{a}}_{\mb{n} \rightarrow \mb{n}'}$ are given by
\begin{equation}
\delta^{(1)}\mathbf{x}^{\text{a}}_{\mb{n} \rightarrow \mb{n}'} = - (\tsif - E_{\text{ref}})\left[ \dxf{a} + \mb{y}^{\text{a}*}_{\mb{n}'}(\text{NN + SRBC}) - \mb{y}^{\text{a}*}_{\mb{n}}(\text{NN + SRBC}) \right].
\label{eqn_del_1_ref}
\end{equation}
with energies measured with respect to some constant reference state energy, $E_{\text{ref}}$.

We now recall that $P^{0}_{\mb{n}} W_{\nnpr} = P^{0}_{\mb{n}}\Gamma_{\nnpr}\tau_{\mb{n}}^{-1}$, where $\Gamma_{\nnpr} = \frac{W_{\nnpr}}{\sum_{\mb{n}'} W_{\nnpr} }$ is the probability of a transition from an initial state and $\tau_{\mb{n}} = \frac{1}{\sum_{\mb{n}'} W_{\nnpr}}$ is the mean residence time in state $\mb{n}$. This means $\overline{W}_{ij}$ and $\overline{b}_{i}$ can be empirically estimated from a set $\mathcal{D}$ of single-step kinetic Monte Carlo trajectories, $(\mb{n} \rightarrow \mb{n}')$ with the initial states drawn from the equilibrium Boltzmann distribution as
\begin{align}
\overline{W}^{(1)}_{ij}{(\mathcal{D})} = \overline{W}_{ji}^{(\mathcal{D})}  &= \frac{1}{|\mathcal{D}|} \sum_{(\mb{n} \rightarrow \mb{n}') \in \mathcal{D} }  \tau_{\mb{n}}^{-1} \Delta \mb{b}^{\text{a}}_{i}(\nnpr).\Delta \mb{b}^{\text{a}}_{j}(\nnpr) \approx \overline{W}^{(1)}_{ij} = \overline{W}^{(1)}_{ji} \\
\overline{b}^{(1)}_{i}{(\mathcal{D})} &= \frac{1}{|\mathcal{D}|} \sum_{(\mb{n} \rightarrow \mb{n}') \in \mathcal{D} }  \tau_{\mb{n}}^{-1} \delta^{(1)}\mathbf{x}^{\text{a}}_{\mb{n} \rightarrow \mb{n}'} . \Delta \mb{b}^{\text{a}}_{i}(\nnpr) \approx \overline{b}^{(1)}_{i} \label{eqn_b_bar_data}
\end{align}
and used in \Eqn{eqn_lbam_dder} to estimate the optimal $\theta^{{i}*}$.

Now, $\overline{W}^{(1)}_{ij}{(\mathcal{D})}$ is independent of $E_{\text{ref}}$. Therefore, for the optimal solutions $\theta^{i*}$ to also be independent of $E_{\text{ref}}$, $\overline{b}^{(1)}_{i}{(\mathcal{D})}$ must also be independent of $E_{\text{ref}}$. Indeed, using \Eqn{eqn_del_1_ref} and \Eqn{eqn_b_bar_data}, we see that
\begin{align}
\nonumber \overline{b}^{(1)}_{i}{(\mathcal{D})} &= -\frac{1}{|\mathcal{D}|} \sum_{(\mb{n} \rightarrow \mb{n}') \in \mathcal{D} }  \tau_{\mb{n}}^{-1} \left( \tsif \left[ \dxf{a} + \mb{y}^{\text{a}*}_{\mb{n}'}(\text{NN + SRBC}) - \mb{y}^{\text{a}*}_{\mb{n}}(\text{NN + SRBC}) \right] \right) . \Delta \mb{b}^{\text{a}}_{i}(\nnpr) \\
& + \frac{E_{\text{ref}}}{|\mathcal{D}|} \sum_{(\mb{n} \rightarrow \mb{n}') \in \mathcal{D} }  \tau_{\mb{n}}^{-1} \left[ \dxf{a} + \mb{y}^{\text{a}*}_{\mb{n}'}(\text{NN + SRBC}) - \mb{y}^{\text{a}*}_{\mb{n}}(\text{NN + SRBC}) \right] . \Delta \mb{b}^{\text{a}}_{i}(\nnpr)
\end{align}
where, the optimal SRBC parameters $\theta^{i*}_{\text{SRBC}}$ satisfy,
\begin{equation}
\sum_{(\mb{n} \rightarrow \mb{n}') \in \mathcal{D} }  \tau_{\mb{n}}^{-1} \left[ \dxf{a} + \mb{y}^{\text{a}*}_{\mb{n}'}(\text{NN + SRBC}) - \mb{y}^{\text{a}*}_{\mb{n}}(\text{NN + SRBC}) \right] . \Delta \mb{b}^{\text{a}}_{i}(\nnpr)= 0
\end{equation}
when optimized over the dataset $\mathcal{D}$. Thus, using the SRBC basis functions and optimizing for $\theta^{i*}$ on the same data of single-step transitions used to optimize $\theta^{i*}_{\text{SRBC}}$ make the obtained  $\theta^{i*}$ independent of the reference state energy.

\section{Alloy Simulation}
\label{sec_CuNi}
To simulate vacancy diffusion in a Ni-20\%Cu alloy, a dataset of single vacancy jumps occurring from equilibrium states has to be constructed. This was done in a similar procedure as in \cite{Chattopadhyay2024}. The first step is to find initial states from Metropolis Monte Carlo (MC) simulations, and then use Climbing Image Nudged Elastic Band (NEB) calculations to find migration barriers, which are detailed below. All energies were computed using the embedded atom method potential of \cite{Foiles1985}. We also used a constant lattice parameter of $3.539$ \AA \ based on applying Vegard's law to the lattice parameters of pure Cu ($3.615$ \AA) and pure Ni ($3.52$ \AA) obtained from \cite{Foiles1985}. The ASE package \cite{HjorthLarsen2017} and the Onsager package \cite{Trinkle2016} were used to construct supercells, write LAMMPS data and for symmetry analysis.
\subsection{Generating Initial States}
\label{sec_CuNi_MC}
\begin{figure}
     \centering
     \includegraphics[width=\wholefigwidth]{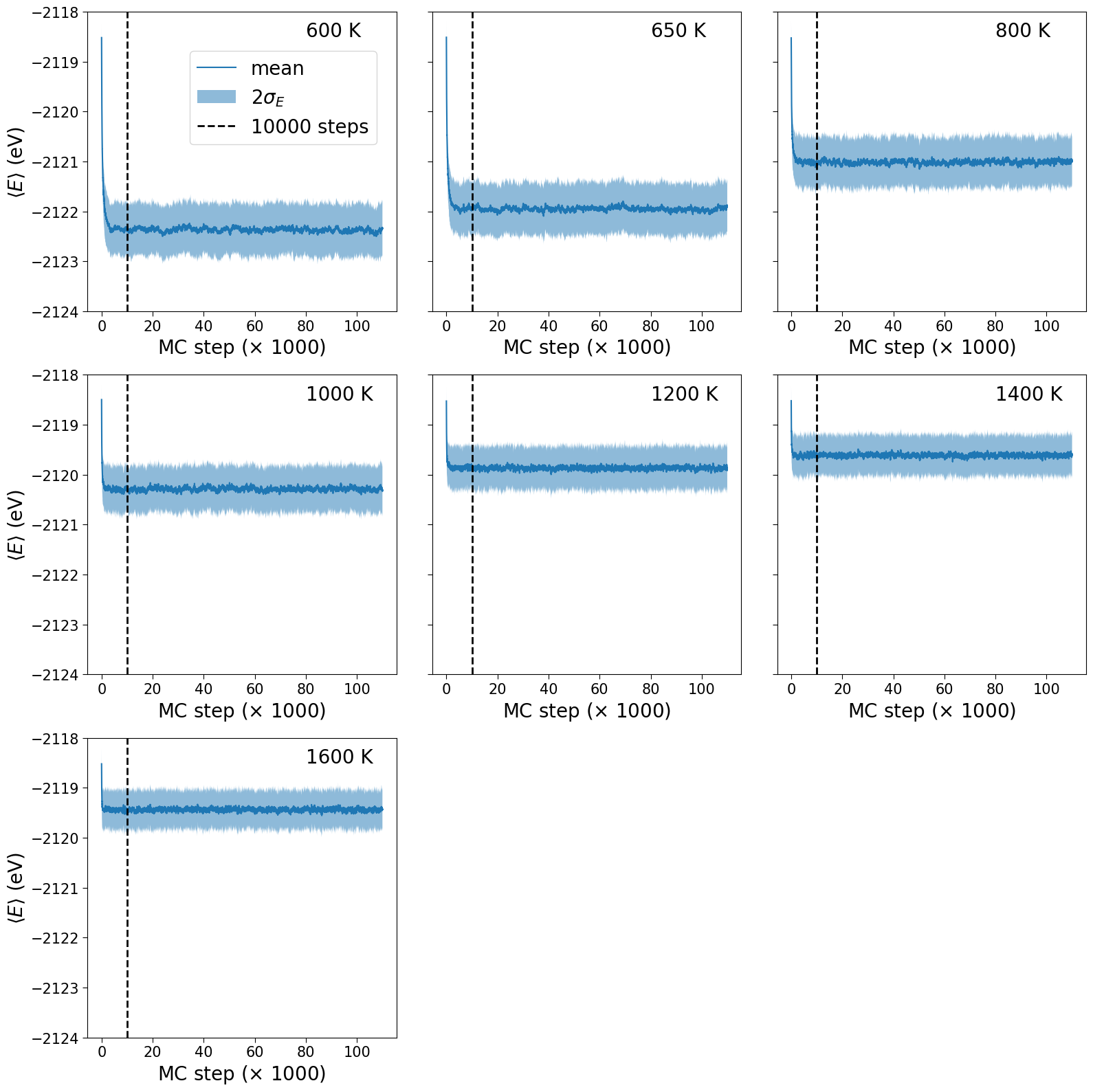}
     \caption{
	Energy evolution of the Ni-20\%Cu alloy during MC simulation, averaged across 200 independent runs. The shaded region corresponds to two standard deviations ($\pm \sigma_{E}$) about the mean value. Energies were seen to be sufficiently converged after $10^4$ MC steps, shown with the dotted vertical line.
	}
     \label{fig_CuNi_MC}
 \end{figure}

MC simulations to construct the datasets were done using the LAMMPS software for energy optimization \cite{LAMMPS}. All energy relaxations were carried out using damped dynamics with the ``ABCFIRE'' algorithm \cite{EcheverriRestrepo2023} up to a maximum force tolerance of $10^{-4}$ eV/\AA \ on any given atom. LAMMPS commands with the parameters used in our energy relaxations are shown below:

{
\begin{verbatim}
min_style fire
timestep 0.001
min_modify norm max abcfire yes tmax 5 dmax 0.01
minimize 0.0 0.0001 100000000 10000000
\end{verbatim}
}
States were gathered from 200 independent simulations. In each such simulation, a state is represented by atomic configurations on a 500-atom cubic FCC supercell, consisting of 1 vacant site, 100 Cu atoms and 399 Ni atoms. Each MC step then consists of attempting to swap the positions of a Ni and Cu atom with each other, with the probability of accepting such a swap being given by $\min[1, \exp(-\beta\delta E)]$ where $\delta E$ is the energy change of the supercell upon performing the swap, calculated from relaxed initial and final configurations. The results are shown in \Fig{fig_CuNi_MC}. We see that on average, $10^4$ MC steps are sufficient to thermalize the states, after which 100 samples were gathered from each at intervals of $10^3$ steps, for a total of $2\times10^4$ states.

\begin{figure}
     \centering
     \includegraphics[width=\wholefigwidth]{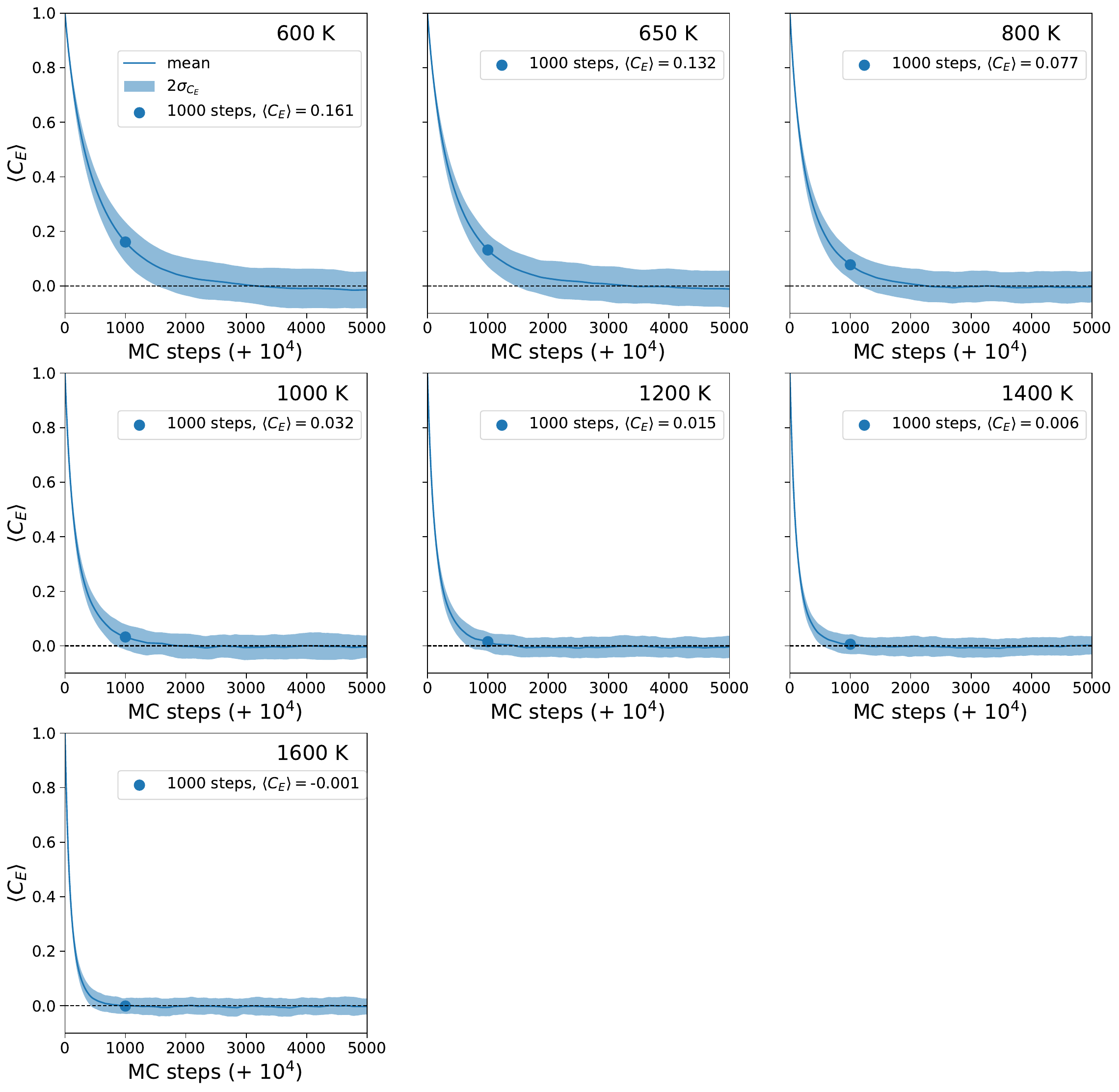}
     \caption{
	Energy autocorrelation function ($C_{E}$) measured as a function of MC steps, starting from sample $10^4$, averaged across 200 independent simulations. At intervals of 1000 MC steps, $C_{E}$ falls below values of 0.2 for 600 K and below 0.1 800 K onwards, indicating sufficiently small correlations between the gathered samples. The shaded region corresponds to $\pm 1$ standard deviation ($\sigma_{C_E}$) above and below the mean.
	}
     \label{fig_CuNi_AC}
 \end{figure}

This interval was selected based on auto-correlation analysis of state energies \cite{Chattopadhyay2024}. These results are in \Fig{fig_CuNi_AC}, showing that on average, gathering samples at intervals of  $10^{3}$ steps leads to sufficiently small correlation between the gathered samples.
 
\subsection{Vacancy Migration Barrier Calculations}
\label{sec_CuNi_NEB}
The LAMMPS parameters used for vacancy migration barrier calculations from the states gathered from the MC simulations are given below.
{
\begin{verbatim}
fix 1 all neb 10.0 parallel neigh
min_style 	 fire
timestep 	 0.001
min_modify 	 norm max abcfire yes tmax 5 dmax 0.01
neb 0.0 0.0001 1000 1000 500 final final_neb.data
# file "final_neb.data" contains the relaxed atom coordinates of
#the final state of the vacancy jump in the appropriate format.
\end{verbatim}
}
To promote equal spacing between images across all NEB calculations, a tangential spring force of $10$ eV/\AA \ was applied to the images.

Prior to the NEB calculation, both initial and final states (for the 12 vacancy jumps) were relaxed. 11 intermediate images were then used (constructed in LAMMPS using a linear interpolation of the relaxed coordinates from the initial to the final state) to simulate every vacancy jump. NEB calculations were then performed to compute the migration barrier of each jump, with a maximum force tolerance of $10^{-4}$ eV/\AA \ for each atom.

\begin{figure}[t]
     \centering
     \includegraphics[width=\wholefigwidth]{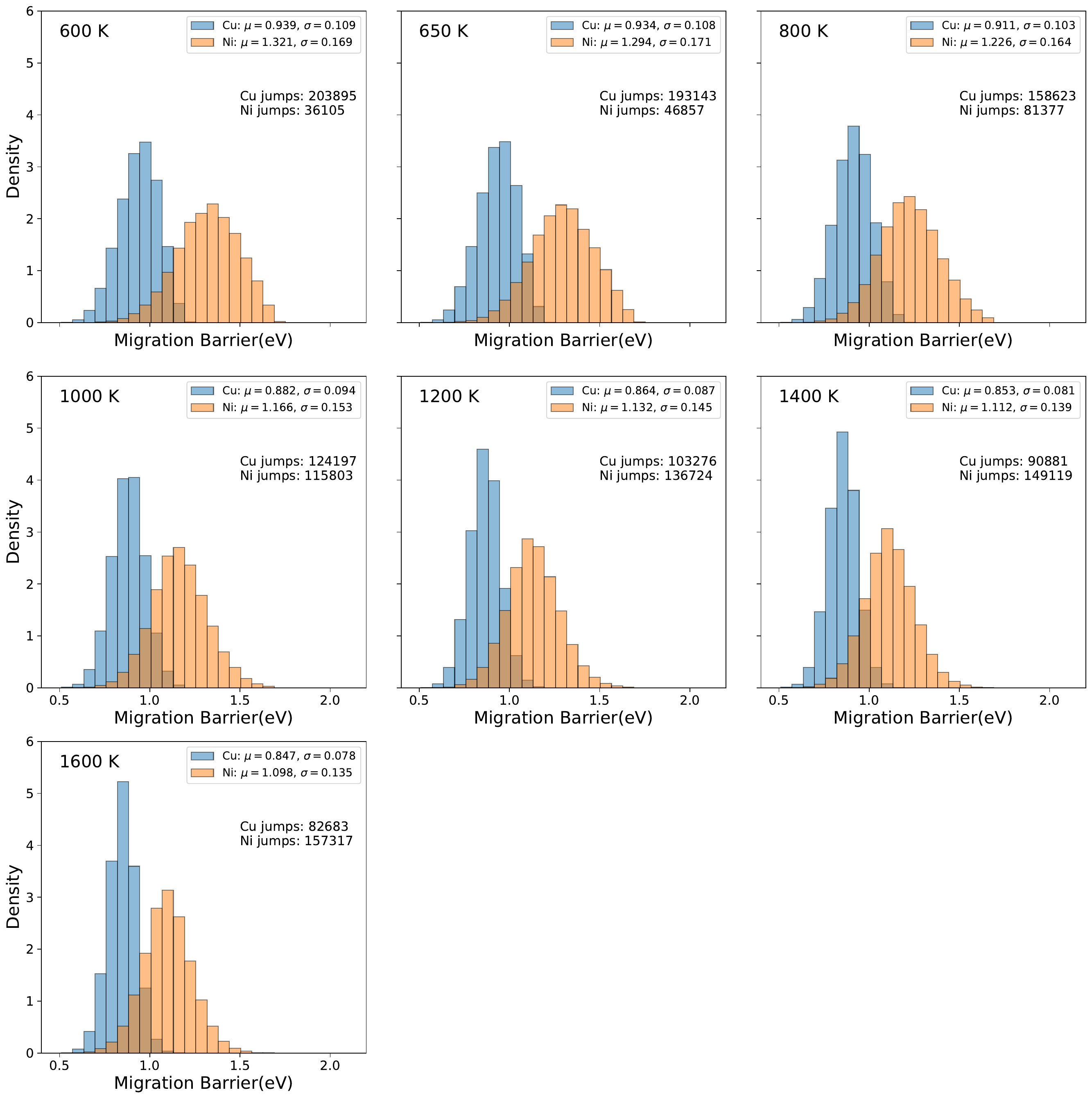}
     \caption{
	Distribution of energy barriers out of the initial states in the Ni-20\%Cu alloy at different temperatures. The total number of Cu and Ni jumps existing around the vacancy in all the equilibrium states are also indicated in the plots. At low temperatures, the vacancy shows a strong preference for Cu atoms around it, while increasing disorder with higher temperatures causes the number of possible Ni jumps to also increase.
	}
     \label{fig_barrier_stats}
 \end{figure}

The distribution of vacancy migration barriers across all simulated temperatures is shown in \Fig{fig_barrier_stats} for the equilibrium initial states drawn from the MC simulations. In general, vacancy-Cu jumps have smaller barriers than vacancy-Ni jumps. However, as the temperature increases, more Ni atoms surround the vacancy (as observed by their growing number of observed jumps) and correspondingly there is also a reduction in the barriers of the jumps of both atoms. This is in agreement with our observation in the main text that the vacancy has a tendency to be surrounded by Cu atoms.

\subsection{Machine Learning Prediction of Vacancy Diffusivity}
\label{sec_CuNi_NN}
Using the NEB details mentioned above, single-step mean residence time kinetic Monte Carlo simulations (as described in \cite{Chattopadhyay2024}) were performed on the states gathered from the MC simulations to generate datasets of $2 \times 10^4$ single vacancy jumps. Migration barriers were then also computed from the final states of these vacancy jumps to allow SRBC corrections. Of the $2\times10^4$ such transitions, $10^4$ were used for training neural network models to minimize the transport coefficient following \Eqn{eqn_var_pri}, with the other $10^4$ used as a validation set. It was ensured that initial states gathered from the same MC simulation run do not enter both the training and validation sets.
 
The details of the neural network design and implementation can be found in \cite{Chattopadhyay2024}, but very briefly, we use the space-group symmetry preserving convolution neural network design of \cite{Cohen2016}, with the last network layer adapted to predict symmetry-equivariant vector-valued quantities for an input state. The neural networks had the same size (i.e, number of parameters) as the optimal networks found for model binary systems in \cite{Chattopadhyay2024}. We used the PyTorch package \cite{paszke2019} to implement and train our neural networks, using the ADAM algorithm \cite{kingma2017} for gradient descent with a learning rate of $10^{-3}$.
\begin{figure}
     \centering
     \includegraphics[width=\wholefigwidth]{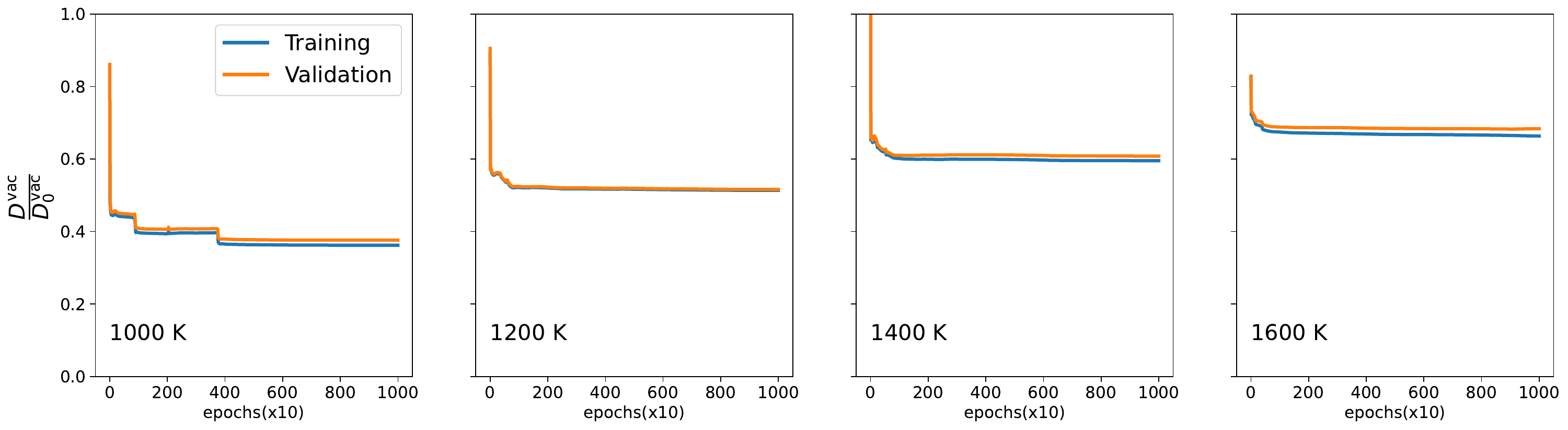}
     \caption{
	Neural network training curves to predict vacancy diffusion coefficients $D^{\text{v}}$ in the Ni-20\%Cu alloy, shown in units of $D^{\text{v}}_{0}$, obtained by setting $\pmb{\eta}^{\text{v}} = \mb{0}$. All networks were trained to a maximum of $10^4$ epochs, and were found to converge quite well by that time. The training and validation sets show very good agreement, indicating a good sampling of the configurational space.
	}
     \label{fig_NN_evo}
 \end{figure}
The results of the neural network training are shown in \Fig{fig_NN_evo}, in which $D^{\text{v}}_{0}$ is the bare single-step KMC diffusion coefficient. All networks were found to converge within $10^4$ epochs, and there is very good agreement between the training and validation sets.  
As shown in \Fig{fig_NN_SRBC}, we found that the neural network at 1000 K can be used to predict the vacancy diffusion coefficient at other temperatures as well. As such, we used the 1000 K network to also predict diffusion coefficients at lower temperatures. Once the neural networks were trained, we then used the SRBC method to correct the relaxation vectors. As seen from \Fig{fig_NN_SRBC}, the resulting transport coefficients cover a wide range of correlation factors, from below 0.1 to above 0.6.
\begin{figure}
     \centering
     \includegraphics[width=4in]{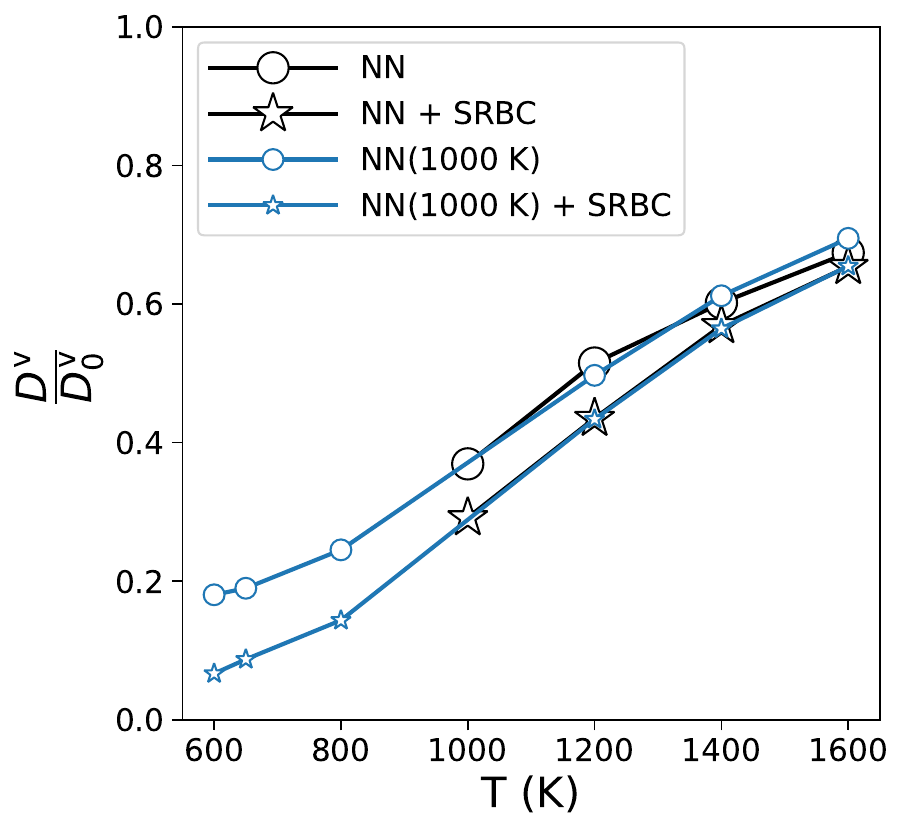}
     \caption{
	Predicted vacancy diffusivity, $D^{\text{v}}$ shown in units of $D^{\text{v}}_{0}$. The neural networks (NNs) are able to substantially capture the correlation effects in the system, with subsequent SRBC correction providing more accurate results. All samples (both training and validation sets) were considered in these plots as they were found to be in reasonable agreement with each other (see \Fig{fig_NN_evo}).
	}
     \label{fig_NN_SRBC}
 \end{figure}

\subsection{Relaxation Vector Change with Temperature}
\label{sec_rel_temp_sens}
Statistics of the approximated values of $\frac{d\pmb{\eta}^{\text{v}}}{d\beta}$ for the Ni-20\%Cu alloy are shown in \Tab{tab_dbeta_rel_norms}. The approximations $\frac{d \mathbf{y}^{\text{v}*}_{\mb{n}}\text{(NN + SRBC)}}{d\beta} \approx \frac{d\pmb{\eta}^{\text{v}}_{\mb{n}}}{d\beta}$ described in \Sec{sec_dbeta_y_SRBC} are labelled the ``direct'' approximations, while the linear basis approximations of \Eqn{eqn_LBAM_dy} in \Sec{sec_LBAM} are labelled the ``variational'' approximations.
\begin{table}[!h]
\centering
\caption{
Statistics (mean $\pm$ standard deviation) for the magnitudes of the approximations for $\frac{d\pmb{\eta}^{\text{v}}}{d\beta}$ for the Ni-20\%Cu alloy, in \AA-eV units (including both initial and final states of a vacancy jump). The ``direct'' approximation is obtained as described in  \Sec{sec_dbeta_y_SRBC}, while the ``variational'' approximations are described in \Sec{sec_LBAM}.
}
\begin{footnotesize}
\begin{tabular}{c|c|c}
&  Direct & Variational\\
\hline \hline
600 K & $0.012 \pm 0.007$ & $0.008 \pm 0.006$\\
650 K & $0.013 \pm 0.007$ & $0.008 \pm 0.005$\\
800 K & $0.016 \pm 0.010$ & $0.004 \pm 0.004$\\
1000 K & $0.031 \pm 0.026$ & $0.026 \pm 0.019$\\
1200 K & $0.066 \pm 0.046$ & $0.062 \pm 0.036$\\
1400 K & $0.087 \pm 0.053$ & $0.088 \pm 0.038$\\
1600 K & $0.102 \pm 0.057$ & $0.096 \pm 0.039$\\
\hline \hline
\end{tabular}
\end{footnotesize}
\label{tab_dbeta_rel_norms}
\end{table}
 We find that overall, both the ``direct'' and ``variational'' approaches show good agreement with each other. At low temperatures (800 K and below), some relatively higher disagreement is observed, particularly at 800 K, but the values themselves are much smaller. Moreover, as seen in the main text, they do not differ much in their influence on $\frac{dQ^{\text{v}}}{d\beta}$.
 
To see how much the relaxation vectors $\pmb{\eta}^{\text{v}}$ vary across temperatures, we compute relaxation vector approximations at different temperatures $T$, for states taken from a given dataset as follows:
\begin{equation}
y^{\text{v}*}_{\mb{n}}\text{(NN + SRBC)}(T) = \mb{y}^{\text{v}*}_{\mb{n}'}(\text{NN}) + \sum_{i} \theta^{i*}_{\text{SRBC}}(T) \mb{b}^{\text{v}}_{i}({\mb{n}}, T) \approx \pmb{\eta}^{\text{v}}_{\mb{n}} (T)
\label{eqn_temp_var_rel}
\end{equation}
where $\theta^{i*}_{\text{SRBC}}(T)$ are the SRBC parameters as obtained by optimizing $D^{\text{v}}$ at each temperature in \Fig{fig_NN_SRBC}, i.e., using the dataset constructed at that temperature. The basis vectors $\mb{b}^{\text{v}}_{i}({\mb{n}}, T)$ are the same as in \Eqn{eqn_basis_srbc} (with ``a'' = ``v'' for the vacancy) but with the jump probabilities recomputed at temperature $T$. $\mb{y}^{\text{v}*}_{\mb{n}'}(\text{NN})$ are the optimal neural network predictions for $\pmb{\eta}^{\text{v}}$, which are temperature-independent. The results are shown in \Tab{tab_rel_vec_norms}.

\begin{table}[!h]
\centering
\caption{
Vacancy relaxation vector magnitudes (in \AA) of states across different temperatures in the Ni-20\%Cu alloy. Each row corresponds to the statistics (mean $\pm$ standard deviation) obtained with the dataset of states constructed at the corresponding indicated temperature (including both initial and final states of a vacancy jump), while each column corresponds to the temperature at which the relaxation vectors were evaluated using the corresponding optimized SRBC coefficients using \Eqn{eqn_temp_var_rel}.
}
\begin{footnotesize}
\begin{tabular}{c|c|c|c|c|c|c|c}
&  $\theta^{i*}_{\text{SRBC}}$(600 K) & $\theta^{i*}_{\text{SRBC}}$(650 K) & $\theta^{i*}_{\text{SRBC}}$(800 K) & $\theta^{i*}_{\text{SRBC}}$(1000 K) & $\theta^{i*}_{\text{SRBC}}$(1200 K) & $\theta^{i*}_{\text{SRBC}}$(1400 K) & $\theta^{i*}_{\text{SRBC}}$(1600 K) \\
\hline \hline
600 K &  $2.166 \pm 0.980$ & $2.165 \pm 0.978$ & $2.190 \pm 0.969$ & $2.199 \pm 0.964$ & $2.211 \pm 0.948$ & $2.263 \pm 0.959$ & $2.247 \pm 0.968$ \\
650 K &  $2.143 \pm 0.964$ & $2.141 \pm 0.962$ & $2.160 \pm 0.952$ & $2.164 \pm 0.945$ & $2.169 \pm 0.925$ & $2.215 \pm 0.935$ & $2.199 \pm 0.944$ \\
800 K &  $2.117 \pm 0.939$ & $2.113 \pm 0.939$ & $2.117 \pm 0.928$ & $2.101 \pm 0.915$ & $2.082 \pm 0.892$ & $2.110 \pm 0.895$ & $2.089 \pm 0.897$ \\
1000 K &  $2.066 \pm 0.905$ & $2.064 \pm 0.910$ & $2.055 \pm 0.896$ & $2.017 \pm 0.877$ & $1.981 \pm 0.852$ & $1.988 \pm 0.849$ & $1.954 \pm 0.841$ \\
1200 K &  $2.127 \pm 0.902$ & $2.128 \pm 0.910$ & $2.099 \pm 0.900$ & $2.044 \pm 0.878$ & $1.989 \pm 0.857$ & $1.977 \pm 0.848$ & $1.939 \pm 0.830$ \\
1400 K &  $1.801 \pm 0.763$ & $1.804 \pm 0.777$ & $1.782 \pm 0.759$ & $1.717 \pm 0.733$ & $1.669 \pm 0.707$ & $1.652 \pm 0.696$ & $1.587 \pm 0.673$ \\
1600 K &  $1.858 \pm 0.790$ & $1.867 \pm 0.807$ & $1.834 \pm 0.787$ & $1.765 \pm 0.760$ & $1.713 \pm 0.735$ & $1.690 \pm 0.721$ & $1.625 \pm 0.696$ \\
\hline \hline
\end{tabular}
\end{footnotesize}
\label{tab_rel_vec_norms}
\end{table}
We find that the temperature-sensitivity of the relaxation vectors, particularly at lower temperatures can be considered small, in agreement with \Tab{tab_dbeta_rel_norms}, but the states at the high temperature datasets (1000 K onward) are more sensitive. From $600$ K to $1600$ K, relative to the values at $600$ K, the average relaxation vector magnitudes of the states in the datasets at $600$ K, $650$ K, $800$ K, $1000$ K, $1200$ K, $1400$ K and $1600$ K change by $3.740$ \%, $2.613$ \%, $1.323$ \%, $5.421$ \%, $8.839$ \%, $11.882$ \%, $12.540$ \% respectively, with the corresponding absolute changes being $0.081$ \AA, $0.056$ \AA, $0.028$ \AA, $0.112$ \AA, $0.188$ \AA, $0.214$ \AA and $0.233$ \AA. This is consistent with \Tab{tab_dbeta_rel_norms} that the approximated magnitude of $\frac{d\pmb{\eta}^{\text{v}}}{d\beta}$ also generally increases for states from higher temperature datasets.

\section{Relaxation Derivatives with Respect to Macroscopic Variables}
\label{sec_general}
The approach of \Sec{sec_var_rel_der} can also be extended to more general cases. We still assume that the transition rates $W_{\nnpr}$ follow an Arrhenius form, but consider all energies, pre-factors and displacements to be functions of some external macroscopic variable  $\lambda$ (such as strain, applied electric/magnetic fields etc) whose analytical forms are assumed to be known. In such a situation, the state probabilities, transition rates and displacements can be generally written as
\begin{align}
&P^{0}_{\mb{n}} = \exp\left[-\beta F_{\mb{n}}(\beta, \lambda)\right]\\
&W_{\nnpr} = \exp\left(-\beta \left[F^{TS}_{\mb{n}, \mb{n}'}(\beta, \lambda) - F_{\mb{n}}(\beta, \lambda)\right] \right) \ \ \text{(inverse time units)}\\
&\dxa(\beta, \lambda) = - \dxarev(\beta, \lambda)
\end{align}
where all energy and entropy terms are absorbed into $F(\beta, \lambda)$. We still assume that $F^{TS}_{\mb{n}, \mb{n}'}(\beta, \lambda) = F^{TS}_{\mb{n}', \mb{n}}(\beta, \lambda)$ is a transition state property to preserve detailed balance, and assume that the diffusion network (state space and transitions) remains unchanged under the applied external variable $\lambda$, which acts to only change the energies and displacements (for example, we do not consider any phase transformation under applied external fields that can change the state space).

\Eqn{eqn_rel_poiss} can then be written as
\begin{equation}
\sum_{\mb{n}'}\exp\left[-\beta F^{TS}_{\mb{n}, \mb{n}'}(\beta, \lambda) \right]\left[\ef(\beta, \lambda) - \ei(\beta, \lambda)\right] = -\sum_{\mb{n}'}\exp\left[-\beta F^{TS}_{\mb{n}, \mb{n}'}(\beta, \lambda) \right]\dxa(\beta, \lambda)
\label{eqn_poiss_general}
\end{equation}

Assuming that all functions of $\lambda$ are differentiable $m$ times, using the general Leibniz rule, we can write
\begin{align}
&\nonumber\sum_{\mb{n}'} \sum_{i=0}^{m} {m \choose i} \left(\frac{d^{i}\exp\left[-\beta F^{TS}_{\mb{n}, \mb{n}'}(\beta, \lambda) \right]}{d\lambda^{i}}\right) \left(\frac{d^{m-i}\ef(\beta, \lambda)}{d\lambda^{m-i}} - \frac{d^{m-i}\ei(\beta, \lambda)}{d\lambda^{m-i}}\right)\\
&= -\sum_{\mb{n}'} \sum_{i=0}^{m} {m \choose i} \left(\frac{ d^{i}\exp\left[-\beta F^{TS}_{\mb{n}, \mb{n}'}(\beta, \lambda) \right]}{d\lambda^{i}}\right) \left(\frac{d^{m-i}\dxa(\beta, \lambda)}{d\lambda^{m-i}}\right)
\end{align}
We separate out the term $i=0$ in the sums on both sides to get
\begin{align}
\nonumber&\sum_{\mb{n}'} \exp\left[-\beta F^{TS}_{\mb{n}, \mb{n}'}(\beta, \lambda) \right] \left(\frac{d^{m}\ef(\beta, \lambda)}{d\lambda^{m}} - \frac{d^{m}\ei(\beta, \lambda)}{d\lambda^{m}}\right)\\
\nonumber&+ \sum_{\mb{n}'} \sum_{i=1}^{m} {m \choose i} \left(\frac{d^{i}\exp\left[-\beta F^{TS}_{\mb{n}, \mb{n}'}(\beta, \lambda) \right]}{d\lambda^{i}}\right) \left(\frac{d^{m-i}\ef(\beta, \lambda)}{d\lambda^{m-i}} - \frac{d^{m-i}\ei(\beta, \lambda)}{d\lambda^{m-i}}\right)\\
\nonumber&= -\sum_{\mb{n}'}\exp\left[-\beta F^{TS}_{\mb{n}, \mb{n}'}(\beta, \lambda) \right] \left(\frac{d^{m}\dxa(\beta, \lambda)}{d\lambda^{m}}\right)\\
& -\sum_{\mb{n}'} \sum_{i=1}^{m}{m \choose i} \left(\frac{ d^{i}\exp\left[-\beta F^{TS}_{\mb{n}, \mb{n}'}(\beta, \lambda) \right]}{d\lambda^{i}}\right) \left(\frac{d^{m-i}\dxa(\beta, \lambda)}{d\lambda^{m-i}}\right)
\end{align}
Or,
\begin{align}
\nonumber&\sum_{\mb{n}'} \exp\left[-\beta F^{TS}_{\mb{n}, \mb{n}'}(\beta, \lambda) \right] \left(\frac{d^{m}\ef(\beta, \lambda)}{d\lambda^{m}} - \frac{d^{m}\ei(\beta, \lambda)}{d\lambda^{m}}\right)\\
\nonumber&= -\sum_{\mb{n}'}\exp\left[-\beta F^{TS}_{\mb{n}, \mb{n}'}(\beta, \lambda) \right] \left(\frac{d^{m}\dxa(\beta, \lambda)}{d\lambda^{m}}\right)\\
& -\sum_{\mb{n}'} \sum_{i=1}^{m}{m \choose i} \left(\frac{ d^{i}\exp\left[-\beta F^{TS}_{\mb{n}, \mb{n}'}(\beta, \lambda) \right]}{d\lambda^{i}}\right) \frac{d^{m-i}}{d\lambda^{m-i}}\left[\dxa(\beta, \lambda) + \ef(\beta, \lambda) - \ei(\beta, \lambda) \right]
\end{align}
Then we note that,
\begin{equation}
\frac{d^{i}\exp\left[-\beta F^{TS}_{\mb{n}, \mb{n}'}(\beta, \lambda) \right]}{d\lambda^{i}} =
\exp\left[-\beta F^{TS}_{\mb{n}, \mb{n}'}(\beta, \lambda) \right]g^{i}_{\lambda}\left[ -\beta F^{TS}_{\mb{n}, \mb{n}'}(\beta, \lambda) \right]
\end{equation}
where $g^{i}_{\lambda}(x)$ is a function of the first $i$ derivatives of $x$ with respect to $\lambda$ , which may be computed using methods such as the Faa di Bruno formula \cite{Constantine1996}.
We then have
\begin{align}
\nonumber&\sum_{\mb{n}'} \exp\left[-\beta F^{TS}_{\mb{n}, \mb{n}'}(\beta, \lambda) \right] \left(\frac{d^{m}\ef(\beta, \lambda)}{d\lambda^{m}} - \frac{d^{m}\ei(\beta, \lambda)}{d\lambda^{m}}\right)\\
\nonumber&= -\sum_{\mb{n}'}\exp\left[-\beta F^{TS}_{\mb{n}, \mb{n}'}(\beta, \lambda) \right] \left(\frac{d^{m}\dxa(\beta, \lambda)}{d\lambda^{m}}\right)\\
& -\sum_{\mb{n}'} \exp\left[-\beta F^{TS}_{\mb{n}, \mb{n}'}(\beta, \lambda) \right] \sum_{i=1}^{m}{m \choose i} g^{i}_{\lambda}\left[ -\beta F^{TS}_{\mb{n}, \mb{n}'}(\beta, \lambda) \right]  \frac{d^{m-i}}{d\lambda^{m-i}}\left[\dxa(\beta, \lambda) + \ef(\beta, \lambda) - \ei(\beta, \lambda) \right]
\end{align}
Or,
\begin{align}
\nonumber&\sum_{\mb{n}'} \exp\left[-\beta F^{TS}_{\mb{n}, \mb{n}'}(\beta, \lambda) \right] \left(\frac{d^{m}\ef(\beta, \lambda)}{d\lambda^{m}} - \frac{d^{m}\ei(\beta, \lambda)}{d\lambda^{m}}\right)\\
\nonumber&= -\sum_{\mb{n}'}\exp\left[-\beta F^{TS}_{\mb{n}, \mb{n}'}(\beta, \lambda) \right] \left[ \left(\frac{d^{m}\dxa(\beta, \lambda)}{d\lambda^{m}}\right)\right.\\
&+ \left.\sum_{i=1}^{m}{m \choose i} g^{i}_{\lambda}\left[ -\beta F^{TS}_{\mb{n}, \mb{n}'}(\beta, \lambda) \right]  \frac{d^{m-i}}{d\lambda^{m-i}}\left[\dxa(\beta, \lambda) + \ef(\beta, \lambda) - \ei(\beta, \lambda) \right] \right]
\end{align}
Then, writing
\begin{align}
\nonumber\ddx{m}{a}(\beta, \lambda) &= \left(\frac{d^{m}\dxa(\beta, \lambda)}{d\lambda^{m}}\right)\\
&+ \sum_{i=1}^{m}{m \choose i} g^{i}_{\lambda}\left[ -\beta F^{TS}_{\mb{n}, \mb{n}'}(\beta, \lambda) \right]  \frac{d^{m-i}}{d\lambda^{m-i}}\left[\dxa(\beta, \lambda) + \ef(\beta, \lambda) - \ei(\beta, \lambda) \right] \label{eqn_dx_m},
\end{align}
and noting that $\ddx{m}{a}(\beta, \lambda) = -\ddr{m}{a}(\beta, \lambda)$ we get
\begin{align}
\sum_{\mb{n}'} \exp\left[-\beta F^{TS}_{\mb{n}, \mb{n}'}(\beta, \lambda) \right] \left(\frac{d^{m}\ef(\beta, \lambda)}{d\lambda^{m}} - \frac{d^{m}\ei(\beta, \lambda)}{d\lambda^{m}}\right)
= -\sum_{\mb{n}'}\exp\left[-\beta F^{TS}_{\mb{n}, \mb{n}'}(\beta, \lambda) \right] \ddx{m}{a}(\beta, \lambda)
\end{align}
Or, multiplying both sides with $\exp\left[\beta F_{\mb{n}}(\beta, \lambda)\right]$,
\begin{align}
\sum_{\mb{n}'} W_{\nnpr} \left(\frac{d^{m}\ef(\beta, \lambda)}{d\lambda^{m}} - \frac{d^{m}\ei(\beta, \lambda)}{d\lambda^{m}}\right)
= -\sum_{\mb{n}'}W_{\nnpr} \ddx{m}{a}(\beta, \lambda)
\end{align}
From which, the derivatives $\frac{d^{m}\ei(\beta, \lambda)}{d\lambda^{m}}$ can be solved as
\begin{equation}
\frac{d^{m}\ei(\beta, \lambda)}{d\lambda^{m}} = \text{argmin}_{y^{(m)}_{\mb{n}}} \frac{1}{2\dm} \sum_{\mb{n'},\mb{n}''}P^{0}_{\mb{n'}}W_{\mb{n}' \rightarrow \mb{n}''}\left( \ddx{m}{a}(\beta, \lambda) + y^{(m)}_{\mb{n}''} - y^{(m)}_{\mb{n}'}  \right)^2
\end{equation}
provided the analytical forms of all the derivatives in \Eqn{eqn_dx_m} are known and they exist.

\clearpage
\bibliography{References}